\documentclass[floats,prb,twocolumn]{revtex4}
\usepackage{amsmath}
\usepackage{color}
\usepackage{graphicx}
\newcommand{\bk}{{\bf k}}
\newcommand{\bq}{{\bf q}}

\newcommand{\bR}{{\bf R}}

\newcommand{\br}{{\bf r}}

\newcommand{\dprime}{{\prime\prime}}
\newcommand{\YBCO}{$\mathrm{YBa_2Cu_3O_{6+x}}${ }}

\begin{document}
\title{The Effect of Pointlike Impurities on $d_{x^2-y^2}$ Charge Density Waves in Cuprate Superconductors}
\author{W.~A.~Atkinson$^1$\email{billatkinson@trentu.ca}  and
  A.~P.~Kampf$^2$\email{Arno.Kampf@physik.uni-augsburg.de}} \affiliation{{}$^1$Department of Physics and
  Astronomy, Trent University, Peterborough Ontario, Canada, K9J 7B8
  \\ {}$^2$Theoretical Physics III, Center for Electronic Correlations
  and Magnetism, Institute of Physics, University of Augsburg, 86135
  Augsburg, Germany} \date{\today}
\date{\today}
\begin{abstract}
Many cuprate superconductors possess an unusual  charge-ordered phase that is characterized by  an approximate $d_{x^2-y^2}$ intra-unit cell form factor and a finite modulation wavevector $\bq^\ast$.  We study the effects impurities on this charge ordered phase via a single-band model in which bond order is the analogue of charge order in the cuprates.  Impurities are assumed to be pointlike and are treated within the self-consistent t-matrix approximation (SCTMA).   We show that suppression of bond order by impurities occurs through the local disruption of the $d_{x^2-y^2}$ form factor near individual impurities.   Unlike $d$-wave superconductors, where the sensitivity of $T_c$ to impurities can be traced to a vanishing average of the $d_{x^2-y^2}$ order parameter over the Fermi surface,  the response of bond order to impurities is dictated by a few Fermi surface ``hotspots". The bond order transition temperature $T_\mathrm{bo}$ thus follows a different universal dependence on impurity concentration $n_i$ than does the superconducting $T_c$.  In particular, $T_\mathrm{bo}$ decreases more rapidly than $T_c$ with increasing $n_i$  when there is a nonzero Fermi surface curvature at the hotspots.  Based on experimental evidence that the pseudogap is insensitive to Zn doping, we conclude that a direct connection between charge order and the pseudogap is unlikely.  Furthermore, the enhancement of stripe correlations in the La-based cuprates by Zn doping is  evidence that this charge order is also distinct from stripes.
\end{abstract}
\maketitle
\section{Introduction}
Hole-doped cuprate superconductors have a pronounced ``pseudogap" phase, which extends across a large fraction of the phase diagram.  The physical origins of the pseudogap are unsettled, and the recent discovery of charge ordering in the pseudogap phase of a variety of cuprates\cite{Kohsaka:2007hx,Wise:2008,Daou:2010bo,Wu:2011ke,Ghiringhelli:2012bw,Chang:2012vf,Sebastian:2012wh,Barisic:2013kz,Blackburn:2013,Wu:2013,Doiron:2013,Blanco-Canosa:2013,CominScience2014,daSilvaNeto:2014bz,Fujita:2014kg,Huecker:2014vc,Wu:2014vx,daSilvaNeto:2015} has led to questions about a possible relationship between the two.  However, the connection is not straightforward:  while the onset temperature for charge order $T_\mathrm{co}$ coincides with the temperature $T^\ast$ at which the pseudogap opens in single-layer Bi$_2$Sr$_{2-x}$La$_x$CuO$_{6+\delta}$,\cite{CominScience2014} $T_\mathrm{co}$ is substantially smaller than $T^\ast$ in other hole-doped cuprates,\cite{Ghiringhelli:2012bw,Chang:2012vf,Doiron:2013,Wu:2014vx}  and is substantially higher than $T^\ast$ in the electron-doped cuprate Nd$_{2-x}$Ce$_{x}$CuO$_4$.\cite{daSilvaNeto:2015}  It has been argued that some combination of charge, superconducting, and current fluctuations may persist up to $T^\ast$ and could be responsible for the pseudogap in the hole-doped cuprates.\cite{Meier:2014,Hayward:2014,Wang:2014wc,Nie:2014jj,Pepin:2014,Wang:2014it}  Conversely, some experiments appear to indicate that charge order is distinct from the pseudogap\cite{meng:2011prb,CominScience2014} and important aspects of the charge-ordered phase can be explained naturally under the assumption that it grows out of the pseudogap.\cite{Atkinson:2014,Chowdhury:2014b}

The charge order has two distinguishing characteristics.  The first is that it appears to have an approximate ``nematic" or $d_{x^2-y^2}$ form factor, which is most easily understood as a transfer of charge between oxygen sites along the $x$ and $y$ axes in the CuO$_2$ planes.\cite{Fischer:2011,Bulut:2013,Atkinson:2014,Fischer:2014}   The strongest evidence for this comes from tunneling experiments in Bi-based cuprates,\cite{Kohsaka:2007hx,Mesaros:2011s,Fujita:2014kg} and it is further supported by recent x-ray experiments.\cite{Comin:2014vq,Achkar:2014}   It is also noteworthy that a  $d_{x^2-y^2}$ form factor is  widely predicted in calculations.\cite{Metlitski:2010vf,Holder:2012ks,YamasePRB2012,Efetov:2013,Bulut:2013}  

The second characteristic is that the amplitude of the interorbital charge transfer is modulated, with wavevectors $\bq^\ast = (q^\ast,0)$ and $\bq^\ast = (0,q^\ast)$ oriented along the Cu-O bond directions.\cite{Kohsaka:2007hx,Ghiringhelli:2012bw,Chang:2012vf,CominScience2014}   The orientation of  $\bq^\ast$ has been hard to understand theoretically,\cite{Atkinson:2014,Chowdhury:2014b,Wang:2014wc,Chowdhury:2014,Pepin:2014} and in general calculations strongly prefer $\bq$-vectors oriented along the Brillouin zone diagonals.  

The charge order may thus be thought of qualitatively as a ``$d_{x^2-y^2}$ charge density wave''. This charge density wave (CDW) appears to be qualitatively different from the stripe order that has been widely observed in the La-based cuprates La$_{2-x}$Ba$_x$CuO$_4$ and La$_{2-x}$Sr$_x$CuO$_4$.\cite{Thampy:2013ty,Atkinson:2014,Achkar:2014} Perhaps the most compelling distinction is that, whereas stripes in the La-cuprates have static or quasistatic spin and charge modulations whose periods are locked together, there is no apparent correlation between spin and charge degrees of freedom in YBa$_2$Cu$_3$O$_{6+x}$.\cite{Blackburn:2013}

In this work, we address the question of how  a $d_{x^2-y^2}$ CDW responds to  strong-scattering pointlike impurities.  Furthermore, because charge order is known to coexist with superconductivity at low temperatures,\cite{Ghiringhelli:2012bw,Chang:2012vf} we explore the effects of impurities on a mixed superconducting-CDW phase.  We adopt a simplified one-band model in which the analogue of charge order is an anisotropic renormalization of the electron hopping, known  as bond order.  Bond order and superconductivity are driven by a combination of spin exchange and Coulomb interactions between nearest-neighbor lattice sites.\cite{Sau:2013vw}

 While the structure of the CDW will be affected by any preexisting pseudogap,\cite{Atkinson:2014,Chowdhury:2014b} we avoid complications associated with modeling the pseudogap and assume instead that the CDW grows out of the full Fermi surface.  The impurity physics described in this work is sufficiently general, however, that it should equally apply in the pseudogap phase.

We distinguish here between two separate issues.  First, it has been pointed out by several authors\cite{Wang:2014wc,Nie:2014jj} that unidirectional charge density waves break both a $U(1)$ symmetry associated with the location of the CDW and a $Z_2$ symmetry associated with its orientation.  Disorder couples linearly to the CDW, and immediately restores the $U(1)$ symmetry (in the disorder average), leaving an ``electron nematic'' phase that breaks rotational but not translational symmetry.  In the nematic phase, the model then maps onto the random field Ising model, which in three dimensions has a critical disorder strength above which long range rotational order is destroyed.  We note, however, that while long range CDW or nematic order is destroyed, local CDW order persists on a length scale set by the disorder potential.\cite{DelMaestro:2006ke}

The second issue concerns the suppression of the {\em amplitude} of the charge order by impurities.  X-ray scattering experiments on YBa$_2$Cu$_3$O$_{6.6}$ observe a rapid reduction of the charge order by Zn impurities.\cite{Blanco-Canosa:2013}   Naively, this is expected given the $d_{x^2-y^2}$ form factor of the charge ordered state:  any order parameter whose average over the Fermi surface is zero should be rapidly suppressed by isotropic scattering.  This mechanism is responsible, for example, for the well-known breakdown of Anderson's theorem in $d$-wave superconductors,\cite{Schmitt-Rink:1986,Hirschfeld:1988,BalatskyRMP:2006} and as pointed out previously by Ho and Schofield,\cite{Ho:2008hn} a mathematically identical theory describes the suppression of the second order $d_{x^2-y^2}$ Pomeranchuk Fermi surface instability. 

Here, we show that the $d_{x^2-y^2}$ symmetry of the charge order is of marginal importance; rather, there is a rapid suppression of charge order by impurities that can be attributed to the ``hotspot" structure of the charge ordered phase.  In particular, the rate at which charge order is suppressed depends sensitively on the Fermi surface curvature near the hotspots.   Consequently,  the suppression of bond order by impurities follows a different universal relationship than $d$-wave superconductors.   

This is of direct relevance to the cuprates, where Zn substitutes isovalently for Cu and acts as a strong-scattering pointlike impurity.\cite{BalatskyRMP:2006,Hirschfeld:2014,AlloulRMP:2009}  Zn-doping has, in past, been used as an important local probe of the superconducting\cite{BalatskyRMP:2006}  and pseudogap states,\cite{Alloul:1991,Zheng:1993,Zheng:1996wh,AlloulRMP:2009}.  Our work leads us to three main conclusions:  first, we find that charge order is more rapidly suppressed than $d$-wave superconductivity; second, the insensitivity of the pseudogap to Zn doping makes charge order an unlikely cause of the pseudogap; third, the insensitivity of stripes in the La-based cuprates to Zn doping supports that the stripe physics is inherently different from charge order in \YBCO and the Bi-based cuprates.

We begin in Sec.~\ref{sec:meanfield} by introducing the model,  and provide an introductory discussion of how bond order responds to pointlike impurities.  Disorder-averaged equations for the bond order parameter are then derived for the case of a dilute concentration of strong scattering impurities, treated within a self-consistent t-matrix approximation (SCTMA).   

In Sec.~\ref{sec:linearized}, we consider temperatures near the bond order transition temperature $T_\mathrm{bo}$ where the equations can be linearized. These equations are the same for uni-directional and bi-directional (checkerboard) order because the different Fourier components of the bond order decouple near $T_\mathrm{bo}$, and their simplicity allows one to derive approximate analytic expressions for the dependence of $T_\mathrm{bo}$ on the impurity concentration $n_i$.  We find that $T_\mathrm{bo}$ is suppressed more quickly by disorder than is the transition temperature $T_c$ for $d$-wave superconductivity.

In Sec.~\ref{sec:commensurate}, we obtain numerical mean-field solutions for $T < T_\mathrm{bo}$ for situations in which the  bond order is commensurate with a periodicity of $m$ unit cells.  X-ray experiments on YBa$_2$Cu$_3$O$_{6+x}$ generally find two sets of peaks, rotated by 90$^\circ$ relative to each other.  The relative intensity of the peaks is strongly doping dependent\cite{Blanco-Canosa:2014} and in YBa$_2$Cu$_3$O$_{6.54}$, where charge order is strongest,  the peak along the ${\bf a}$-axis is almost undetectably weak.\cite{Blackburn:2013}  This suggests that the two Fourier components of the charge order are not strongly coupled.  Furthermore, analyses of STM experiments in Bi-based cuprates  find nanoscale domains of unidirectional order rather than true biaxial order.\cite{Kohsaka:2007hx,Mesaros:2011s,Fujita:2014kg}  We therefore focus on  unidirectional order, although an extension to multiple Fourier components is conceptually straightforward.   Our calculations lead to a coupled set of equations for the impurity scattering rate (or, more specifically, the self-energy) and the bond order parameters.  Most of the technical details are relegated to the appendices.

Experimentally, the charge ordering temperature $T_\mathrm{co}$ is greater than  $T_c$, and we therefore examine the onset of superconductivity in the presence of bond order in Sec.~\ref{sec:commensurateSC}.  We derive superconducting $T_c$ equations within the SCTMA, which are then solved in conjunction with the self-consistent equations for the bond order parameter.  We find that, in the mixed phase, impurities suppress $T_\mathrm{bo}$ more rapidly than $T_c$, and one may therefore obtain $T_c > T_\mathrm{bo}$ as the impurity concentration increases.

Finally, in Sec.~\ref{sec:discuss}, we discuss our results in the context of Zn-doping experiments in the pseudogap phase, and in the stripe phase of La$_{2-x}$Ba$_x$CuO$_4$.   These experiments show that the pseudogap and stripe phases  respond differently to Zn impurities than the charge ordered phase seen in YBa$_2$Cu$_3$O$_{6+x}$, and therefore likely have a different origin.

\section{Mean-Field Equation for Bond Order}
\label{sec:meanfield}
Following Ref.~\onlinecite{Sau:2013vw},
we adopt a tight-binding model on a square, two-dimensional lattice, representing a single CuO$_2$ plane.  The noninteracting part of the Hamiltonian is 
\begin{equation}
H_0 = \sum_{\bk,\alpha} \epsilon_\bk c^\dagger_{\bk\alpha} c_{\bk\alpha}
\end{equation}
with dispersion $\epsilon_\bk = t_0 - 2t_1(\cos k_x + \cos k_y) + 4t_2 \cos k_x \cos k_y$.  We take $t_1 = 1$, which sets the energy scale for the calculation; in the cuprates the bandwidth $8 t_1$ is of order a few electron volts.  The interacting part of the Hamiltonian contains both nearest-neighbor exchange and Coulomb interactions,
\begin{equation}
H_1 = \frac J8 \sum_{\langle i,j\rangle} \sum_{\alpha,\ldots,\delta} \sum_a \tau^a_{\alpha\beta} \tau^a_{\gamma\delta} c^\dagger_{i\alpha} c_{i\beta} c^\dagger_{j\gamma}c_{j\delta}
+\frac V2 \sum_{\langle i,j\rangle} \hat n_i \hat n_j
\end{equation}
where $\langle i,j\rangle$ refers to nearest-neighbor lattice sites $i$ and $j$ and $\tau^a_{\alpha\beta}$ are Pauli matrices.  We perform a
mean-field decomposition of this interaction in the exchange channel to obtain
\begin{equation}
H_{1\mathrm{bo}} = -\left( \frac{3J}{4} + V \right ) \sum_{\langle i,j\rangle} \sum_{\sigma} \langle c^\dagger_{j\sigma} c_{i\sigma}\rangle c^\dagger_{i\sigma} c_{j\sigma}
\end{equation}
where $ \langle c^\dagger_{j\sigma} c_{i\sigma}\rangle$ is assumed to be independent of the spin $\sigma$.
This term drives the bond-ordering instability, and we therefore define an effective interaction for bond order,
\begin{equation}
J_\mathrm{bo} = \left( \frac{3J}{4} + V \right ).
\end{equation}
A similar decomposition in the particle-particle channel leads to a mean-field superconducting contribution,
\begin{equation}
H_{1\mathrm{sc}} = -J_\mathrm{sc} \sum_{\langle i,j\rangle} \left [\langle c^\dagger_{j\uparrow} c^\dagger_{i\downarrow}\rangle c_{i\downarrow} c_{j\uparrow} + \mathrm{h.c.}\right]
\label{eq:Hsc}
\end{equation}
with
\begin{equation}
J_\mathrm{sc} = \left( \frac{3J}{4} - V \right ).
\end{equation}
While the spin-exchange interaction is attractive in both the bond-order and superconducting channels, the Coulomb interaction enhances bond order and suppresses superconductivity.  This was invoked previously as an explanation for why $T_\mathrm{bo}$ is greater than the superconducting transition temperature $T_ c$.\cite{Sau:2013vw}

We denote the exchange self-energy along the nearest-neighbor bond $i$-$j$ by
\begin{equation}
P_{ji} = -J_\mathrm{bo} \langle c^\dagger_{j\sigma} c_{i\sigma}\rangle.
\label{eq:Pij}
\end{equation}
For illustrative purposes, we solve this self-consistently in real space:  $P_{ji}$ is calculated for each bond by diagonalizing the mean-field Hamiltonian $H_0 + H_{1\mathrm{bo}}$ on an $L\times L$ lattice with periodic boundary conditions; the new $P_{ji}$ are used to update $H_{1\mathrm{bo}}$,  and the process is iterated until self-consistency is achieved.  To obtain a solution, it is necessary to tune the band parameters such that the system size is an integer multiple of the CDW period.

\begin{figure}
\includegraphics[width=\columnwidth]{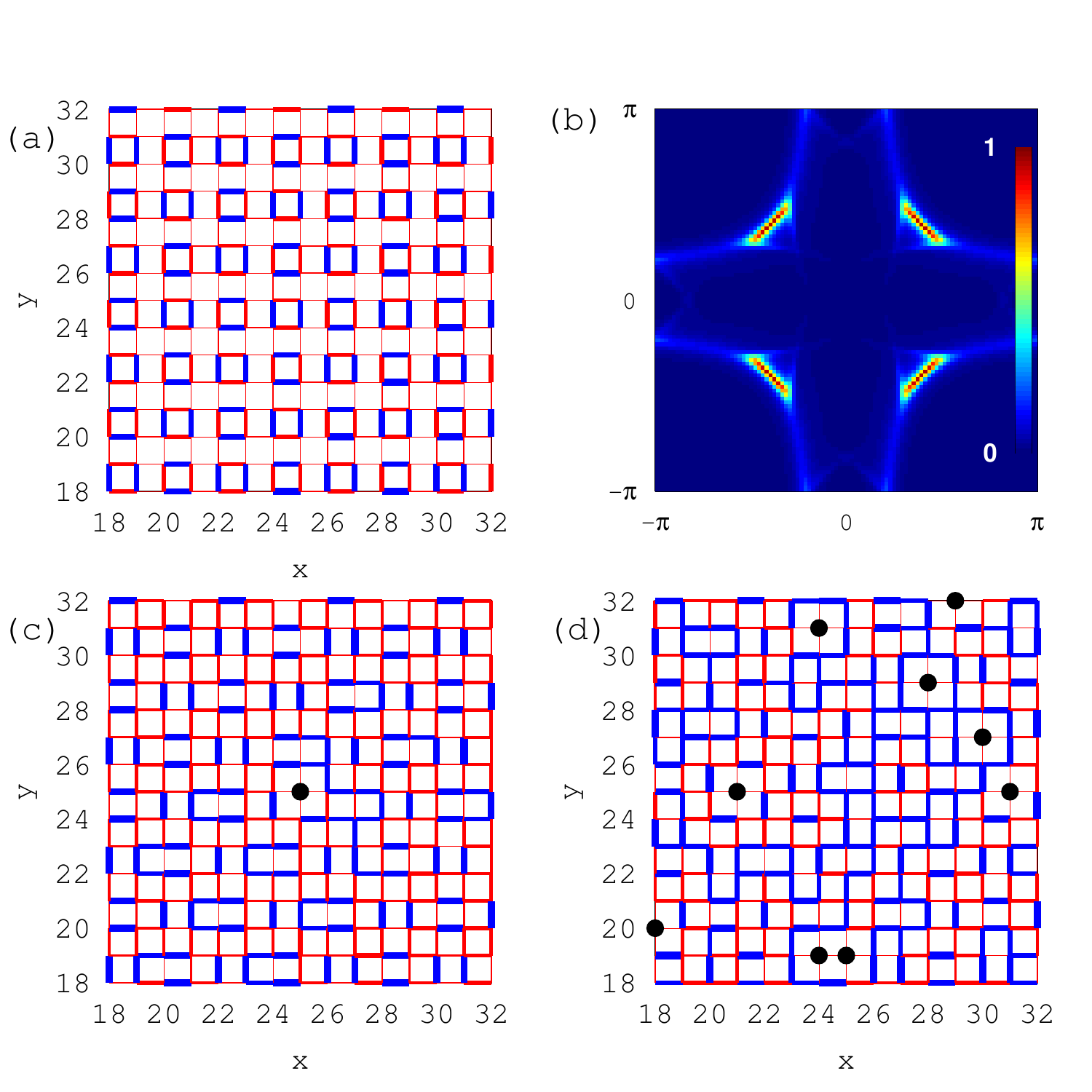}
\caption{Results of self-consistent real-space bond-order calculations on a $48\times48$ lattice. (a) Bond self-energies $P_{ij}$  between nearest-neighbor lattice sites in the clean (impurity-free) limit.  The constant reference self-energy $P_0$ of the homogeneous phase is subtracted, as described in the text, to highlight spatial fluctuations.  Blue (red) line colors indicate that the self-energy is enhanced (reduced) relative to $P_0$, and linewidths indicate the size of the enhancement (reduction).  (b)  $\bk$-dependent spectral function at the Fermi energy for the clean limit. (c) Bond self-energies for a system containing a single impurity, indicated by a black circle at the center of the figure.  Here, $P_0$ has not been subtracted when determining the linewidths; however, the color scheme is the same as in (a).   (c) Bond self-energies for a concentration $n_i = 0.04$ of impurities.  Impurity locations are indicated by black circles.
 Model parameters are  $t_0 = 0.85$, $t_1 = 1.0$, $t_2 = -0.5$,  $J_\mathrm{bo} = 2.40$, $V_i=10.0$, and $T=0.25$.}
\label{fig:BdG}
\end{figure}

Typical results for a clean lattice are shown in Fig.~\ref{fig:BdG}(a). To highlight the bond ordering, we have subracted off the uniform exchange self-energy $P_0$ of the homogeneous phase, which is obtained by requiring all bonds to be equivalent in the self-consistent calculation.  The amplitudes and signs of the shifts in the bond self-energy, relative to $P_0$, are shown by the thicknesses and colors of the lines connecting nearest-neighbor sites.  For the parameters chosen, the modulation amplitude is about 10\% of $P_0$.  The phase shown in Fig.~\ref{fig:BdG}(a) has  Fourier components of equal magitude at  $\bq = \pm(2\pi/4)(1,1)$ and $\bq = \pm(2\pi/4) (1,-1)$, and can be thought of as a $d_{x^2-y^2}$ form factor whose amplitude has a period-4 checkerboard modulation.

The corresponding spectral function $A(\bk,\varepsilon_F)$ at the Fermi energy is shown in Fig.~\ref{fig:BdG}(b).  The bond-order $\bq$-vectors connect segments of Fermi surface near $(\pm \pi,0)$ and $(0,\pm\pi)$, and which points a spectral gap opens.  The residual Fermi surface has a strong intensity along arcs centered on the Brillouin zone diagonals.  Faint residual Fermi surface segments can also be seen along the Brillouin zone boundaries.

Figure~\ref{fig:BdG}(c) shows the effects of adding a single strong-scattering impurity to the lattice, modeled as a potential shift of $V_i =10.0$ at position $(25,25)$.  In this panel, the linewidths indicate the total bond self-energy, including $P_0$; however, the color scale still indicates the shift of $P_{ij}$ relative to $P_0$, as in Fig.~\ref{fig:BdG}(a).  Unsurprisingly, $P_{ij}$ is reduced almost to zero along bonds connecting to the impurity site.  The amplitude of the bond self-energy recovers within a lattice spacing of the impurity; however, the modulation pattern is disrupted over a longer length scale.   Thus, suppression of bond order does not simply imply that the bond self-energies vanish, but rather that the form factor is disrupted near impurities.  This is evident in Fig.~\ref{fig:BdG}(d) where the structure shown in (a) is entirely disrupted by a 4\% impurity concentration.  This scenario is to be contrasted with that presented in, for example, Ref.~\onlinecite{Nie:2014jj} where smooth impurity potentials disrupt long range order but leave the form factor locally intact.

To proceed further, we study the disorder-averaged mean-field equations for bond order, and subsequently  superconductivity, for a dilute concentration of strong scattering impurities.  Because disorder-averaging restores translational symmetry, it is useful to Fourier transform Eq.~(\ref{eq:Pij}) to $\bk$-space,
\begin{equation}
H_{1\mathrm{bo}} = \sum_{\bk,\bq,\sigma} P_\bk(\bq) c^\dagger_{\bk+\bq \sigma}c_{\bk\sigma}
\label{eq:Hbo}
\end{equation}
where
\begin{eqnarray}
P_\bk(\bq) &=& \frac{1}{N} \sum_{i,\delta} P_{i+\delta,i} e^{i\bk\cdot\delta} e^{-i\bq\cdot \br_i} 
\label{eq:nematic0} \\
&=& - \frac{J_\mathrm{bo}}{N} \sum_{\bk^\prime} 2\left [ \cos (k_x-k_x^\prime) + \cos(k_y-k_y^\prime) \right ] \nonumber \\
&&\times \langle c^\dagger_{\bk^\prime\sigma} c_{\bk^\prime + \bq\sigma} \rangle.
\label{eq:nematic1}
\end{eqnarray}
In this definition of $P_\bk(\bq)$, $\bk$ and $\bk+\bq$ are initial and final momentum labels for electrons that are scattered by the bond order; a common alternative is to take these initial and final points to be at  $\bk-\bq/2$ and $\bk+\bq/2$.  This latter choice is 
less convenient for  systems with commensurate bond order,  which we discuss below. 

Equation~(\ref{eq:nematic1}) is the basic self-consistent equation for $P_\bk(\bq)$.  
It is invariant under $\bk \rightarrow -\bk-\bq$ when $P_\bk(\bq)$ is real, so that $P_\bk(\bq)$ should be even or odd under this transformation.  We then expand Eq.~(\ref{eq:nematic1}) in a set of basis functions that are even or odd under $\bk \rightarrow -\bk-\bq$, via
\begin{equation}
2\left [ \cos (k_x-k_x^\prime) + \cos(k_y-k_y^\prime) \right ] = \sum_{\alpha=1}^4 \eta^\alpha(\bk)
\eta^\alpha(\bk^\prime),
\end{equation}
with
\begin{subequations}
\begin{eqnarray}
\eta^1_\bk &=& \sqrt{2}\cos(k_x+\frac{q_x}{2}),  \\
\eta^2_\bk &=& \sqrt{2}\cos(k_y+\frac{q_y}{2}), \\
\eta^3_\bk &=& \sqrt{2}\sin(k_x+\frac{q_x}{2}),  \\
\eta^4_\bk &=& \sqrt{2} \sin(k_y+\frac{q_y}{2}).
\end{eqnarray}
\label{eq:basis}
\end{subequations}
Then, Eq.~(\ref{eq:nematic1}) reduces to
\begin{equation}
P_\bk(\bq) = \sum_\alpha \eta^\alpha_\bk P^\alpha(\bq), 
\end{equation} 
with
\begin{eqnarray}
P^\alpha(\bq) &=&   -\frac{J_\mathrm{bo} }{N} \sum_{\bk^\prime} \eta^\alpha_{\bk^\prime} \langle c^\dagger_{\bk^\prime\sigma} c_{\bk^\prime +\bq \sigma} \rangle. \\
&=& - \frac{J_\mathrm{bo} }{N} \sum_{\bk^\prime} \eta^\alpha_{\bk^\prime} 
G(\bk^\prime+\bq,\bk^\prime; \tau = 0^-), 
\label{eq:nematicOP}
\end{eqnarray}
where $G(\bk_1,\bk_2;\tau)$ is the Green's function at imaginary times $\tau$ in the presence of bond order.   This equation has an even solution
\begin{equation}
P_\bk(\bq) = P^1(\bq) \eta^1_\bk + P^2(\bq) \eta^2_\bk,
\end{equation}
and an odd solution 
\begin{equation}
P_\bk(\bq) = P^3(\bq) \eta^3_\bk + P^4(\bq) \eta^4_\bk.
\end{equation}
The even solution is the leading instability in all calculations reported here.

Equation~(\ref{eq:nematicOP}) requires an explicit expression for the Green's function, and we consider two cases where closed expressions are possible: (i) temperatures near $T_\mathrm{bo}$ where a linearized Green's function can be obtained and (ii) the case of period-$m$ commensurate order.

\section{Linearized Equations for $T_\mathrm{bo}$}
\label{sec:linearized}
Near the bond ordering transition, $P_\bk(\bq)$ is small, so that $H_{1\mathrm{bo}}$ can be treated perturbatively.  We show in Appendix~\ref{sec:A2} that to linear order in $P_\bk(\bq)$,  Eq.~(\ref{eq:nematicOP}) reduces to a matrix equation for the components $P^\alpha(\bq)$,
\begin{equation}
P^\alpha(\bq) = J_\mathrm{bo}  \sum_\beta F_{\alpha\beta}(\bq) P^\beta(\bq).
\label{eq:linearized}
\end{equation}
Equation (\ref{eq:linearized}) has a solution at the ordering wavevector $\bq^\ast$ when the largest eigenvalue of the matrix $\bf F(\bq^\ast)$ is equal to $1/J_\mathrm{bo}$.   For a given $J_\mathrm{bo}$, we search for the temperature $T_\mathrm{bo}$ at which bond order emerges.  We note that the different $\bq$-vectors  are decoupled at $T_\mathrm{bo}$, with each independently satisfying an equation of the form (\ref{eq:linearized}).  Consequently, the dependence of $T_\mathrm{bo}$ on impurity concentration is the same for uni-directional charge order as it is for bi-directional (checkerboard) order.

\begin{figure}
\includegraphics[width=\columnwidth]{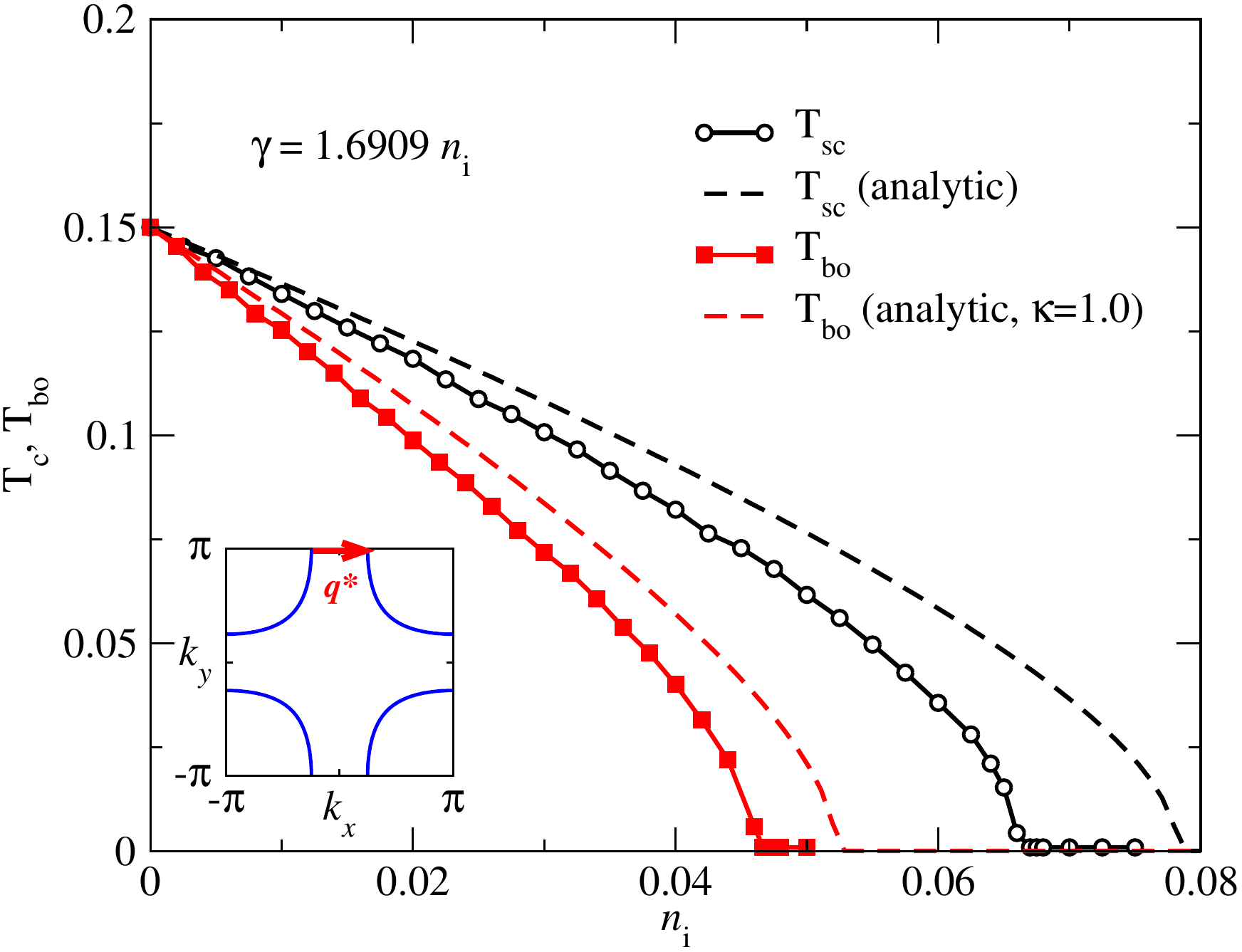}
\caption{Comparison of impurity effects on period-4 axial bond order and $d$-wave superconductivity.  The bond ordering temperature $T_\mathrm{bo}$ and superconducting transition temperature $T_ c$ are shown as a function of the impurity concentration for
impurity potential $V_i  = 10$.    Results are shown for full numerical calculations (solid curves), and for the analytical equations (\ref{eq:digamma}) and (\ref{eq:digammaSC}) (dashed curves).  We have taken $\kappa =1.0$ in Eq.~(\ref{eq:digamma}).   To determine $\gamma$ for the analytical equations, a linear fit was made to $-\mbox{Im }\Sigma_n$ for $n=0$,  as a function of $n_i$: the resulting formula is given in the figure.   {\em Inset:} Fermi surface and $\bq^\ast$.  Band parameters are  $t_0 = 0.85$ and $t_2 = 0.5$.  The interaction strengths $J_\mathrm{bo} = 2.74$ and  $J_\mathrm{sc} = 1.44$  are chosen to obtain $T_\mathrm{bo}^0 = T_ c^0 = 0.15$.}
\label{fig:Tc_vs_Ni}
\end{figure}

A plot of $T_\mathrm{bo}$ versus impurity concentration $n_i$ is shown in Fig.~\ref{fig:Tc_vs_Ni} for period-4 axial order, with $\bq^\ast = (2\pi/4,0)$, for strong scattering impurities.  To obtain this figure, we have tuned the band parameter $t_0$, which controls the filling, such that $\bq^\ast$ connects parallel segments of Fermi surface (see  Fig.~\ref{fig:Tc_vs_Ni} inset).  For comparison, the dependence of  $T_c$ on $n_i$, calculated with the SCTMA (Appendix~\ref{sec:B1}), is shown for a $d$-wave superconductor.  We see that both bond order and superconductivity are suppressed by disorder, but that $T_\mathrm{bo}$ is suppressed more rapidly than $T_c$.

To understand the suppression of  bond order by impurities, we analyze the equations governing $T_\mathrm{bo}$.  The kernel $F_{\alpha\beta}(\bq)$ is (Appendix~\ref{sec:A2})
\begin{equation}
F_{\alpha\beta}(\bq) = -\frac{T}{N} 
\sum_{\bk,n}   
 \frac{ \eta^\alpha_{\bk} (\eta^\beta_{\bk} + {\cal S}^{\bq,\beta}_n) }
 {(i\omega_n - \epsilon_{\bk+\bq} -  \Sigma_n)
 (i\omega_n - \epsilon_{\bk} -  \Sigma_n)},
\label{eq:Fab}
\end{equation}
where $\omega_n = (2n+1)\pi T$ are Matsubara frequencies, $\Sigma_n$ is the impurity  self-energy at $\omega_n$  to zeroth order in $P_\bk(\bq)$, and the linear-order impurity self-energy $\Sigma_n^\bq$ has been factored into components,
\begin{equation}
\Sigma_n^\bq = \sum_\beta {\cal S}_n^{\bq,\beta} P^\beta(\bq).
\end{equation}

For temperatures near $T_\mathrm{bo}$, the SCTMA gives the self-consistent equation for the zeroth order impurity self-energy [Eq.~(\ref{eq:Sigma0})],
\begin{equation}
\Sigma_n = n_i V_i \left [ 1 - \frac{V_i}{N} \sum_\bk \frac{1}{i\omega_n - \epsilon_\bk -\Sigma_n} \right ]^{-1},
\label{eq:Sn}
\end{equation}
where $n_i$ is the impurity concentration and $V_i$ is the impurity potential.   The real part of $\Sigma_n$ acts as a chemical potential shift due to doping by the impurities, and the imaginary part is the negative of the scattering rate $\gamma_n$.   Because $\gamma_n$ has the same sign as $\omega_n$, it behaves qualitatively like a temperature increase:  it can be absorbed into a renormalized Matsubara frequency $\tilde \omega_n = \omega_n +\gamma_n$ whose magnitudes are larger than the unrenormalized frequencies $\omega_n$.
The effect of $\Sigma_n$ is therefore to reduce bond order and suppress $T_\mathrm{bo}$.

The physics of $\Sigma^\bq_n$ is quite different from that of $\Sigma_n$.   We find numerically that when $\Sigma^\bq_n$ is omitted from the self-consistent calculations, $T_\mathrm{bo}$ is reduced.  This is similar to the situation in superconductors, which we review in Appendix~\ref{sec:B1}, where an analogous ``anomalous'' self-energy appears in the equations for $T_c$.  In conventional isotropic $s$-wave superconductors, the $T_c$ enhancement by the anomalous self-energy cancels  the reduction of $T_c$ by $\Sigma_n$ [cf.\ Eq.~(\ref{eq:Tcswave})], consistent with Anderson's statement that $T_c$ is unaffected by disorder.\cite{BalatskyRMP:2006}  The response of $T_c$ to impurities is closely tied to the symmetry of the superconducting order parameter:  in $d$-wave superconductors, the anomalous self-energy vanishes [cf.\ Eq.~(\ref{eq:DSC})] and $T_c$ is strongly reduced by impurities.\cite{Schmitt-Rink:1986,Hirschfeld:1988}

From Eq.~(\ref{eq:SigmaQ2}), the expression for $\Sigma_n^\bq$ is proportional to a weighted average of $P_\bk(\bq)$ over the Brillouin zone:
\begin{equation}
\Sigma^\bq_n \propto  \frac{1}{N} \sum_\bk \frac{P_\bk(\bq)}{(i\omega_n - \epsilon_\bk -\Sigma_n)(i\omega_n - \epsilon_{\bk+\bq} -\Sigma_n)}.
\label{eq:Sqn}
\end{equation}
The sum in Eq.~(\ref{eq:Sqn}) is  weighted towards those points, the so-called ``hotspots'',  for which $\bk$ and $\bk+\bq$ both lie on the Fermi surface.  This has two consequences:  first,  the $\bk$-sum in Eq.~(\ref{eq:Sqn}) does not vanish, even when $P_\bk(\bq)$ has a nominally $d$-symmetric form factor $P_\bk(\bq) \propto \eta^1_\bk - \eta^2_\bk$; second,  $\Sigma^\bq_n$ nonetheless tends to be small because of the limited region of $\bk$-space that contributes to the sum.  Indeed, the omission of $\Sigma^\bq_n$ from the self-consistent calculations only changes $T_\mathrm{bo}$ by a few percent.   We conclude that the sensitivity of $T_\mathrm{bo}$ to impurities is not tied to the symmetry of the form factor, but is a consequence of the central role of hotspots in the $T_\mathrm{bo}$ calculation.

We can integrate Eq.~(\ref{eq:Fab}) analytically under a few simplifying assumptions.  We expand the electronic dispersion around the  Fermi surface hotspots, and ignore the energy dependence of $\Sigma_n$, letting $\Sigma_n \rightarrow -i\gamma\mbox{sgn}(\omega_n)$.  We obtain (see Appendix~\ref{sec:A3}) 
\begin{eqnarray}
\ln \frac{T_\mathrm{bo}^0}{T_\mathrm{bo}} = \frac{1}{\pi} \int_{0}^\pi dk_y \mathrm{Re } \left [
\psi\left( \frac 12 + \frac{\gamma}{2\pi T_\mathrm{bo}} + i \frac{\kappa k_y^2}{2\pi T_\mathrm{bo}} \right ) \right . \nonumber \\
-\left .
\psi\left( \frac 12  + i \frac{\kappa k_y^2}{2\pi T_\mathrm{bo}} \right )
\right ],
\label{eq:digamma}
\end{eqnarray}
where $\kappa$ is the Fermi surface curvature at the hotspots, $\gamma$ is the scattering rate, $T_\mathrm{bo}^0$ is the bond ordering temperature in the clean limit, and $\psi(x)$ is the digamma function.  Equation~(\ref{eq:digamma}) obtains a form similar to the usual result for the  transition temperature of a $d$-wave superconductor, namely\cite{BalatskyRMP:2006}
\begin{equation}
\ln \frac{T_ c^0}{T_ c} = \psi\left ( \frac 12 + \frac{\gamma}{2\pi T_\mathrm{c}} \right )
- \psi\left ( \frac 12 \right ),
\label{eq:digammaSC}
\end{equation}
when the Fermi surface curvature is $\kappa = 0$.  

These analytical expressions are shown in Fig.~\ref{fig:Tc_vs_Ni}. To make the comparison quantitative, we have set $\gamma = -\mbox{Im }\Sigma_{n=0}$,  where $\omega_0$ is the lowest positive Matsubara frequency.  The curvature is
\[
\kappa = \left | \frac{ \partial^2 \epsilon_\bk }{ \partial k_y^2 }\right |_{\bk=(2\pi-\frac{q_x}{2},\pi)}
\]
which is typically a number of order $1$.  It is apparent in Fig.~\ref{fig:Tc_vs_Ni} that the analytical expressions overestimate the transition temperatures somewhat, but that they capture the reduction of $T_\mathrm{bo}$ relative to $T_ c$.   We note that the sensitivity of $T_\mathrm{bo}$ to disorder is {\em in addition} to the reduction of $T_\mathrm{bo}^0$ due to $\kappa$ in the clean limit; indeed, in Fig.~\ref{fig:Tc_vs_Ni} we had to use an inflated value of $J_\mathrm{bo} = 2.74$ relative to the pairing interaction $J_\mathrm{sc} = 1.44$ to obtain $T_\mathrm{bo}^0 = T_ c^0$.

\section{Commensurate Bond Order}
\subsection{Pure bond order}
\label{sec:commensurate}
When the bond order parameter is not small, we can proceed by assuming that the wavevector
is commensurate, with $\bq^\ast = (2\pi/m) (1,0)$ for axial order and $\bq^\ast = 2\pi/m(1,1)$ for diagonal order, where $m$ is an integer.  These describe uni-directional phases, and the extension to bi-directional order is straightforward.  For clarity, we describe only the case of uni-directional order.

\begin{figure}
\includegraphics[width=0.6\columnwidth]{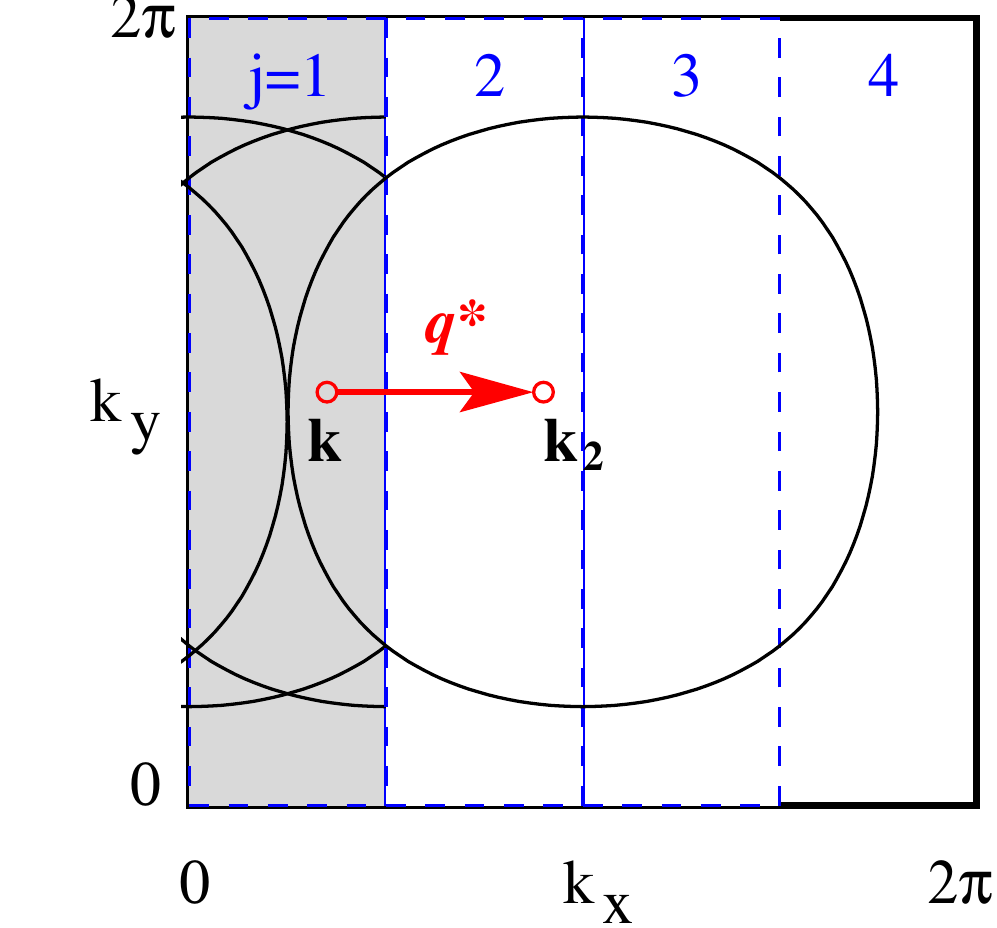}
\caption{Construction of the reduced Brillouin zone.  In this example we consider period-4 axial order ($m=4$), with modulation wavevector $\bq^\ast = (2\pi/4)(1,0)$.  The original Fermi surface (solid black curve) is shown in the full Brillouin zone, which  extends from 0 to $2\pi$ along $k_x$ and $k_y$.  The reduced Brillouin zones have width $|\bq^\ast|$ alond $k_x$, and extend from 0 to $2\pi$ along $k_y$.  The different reduced Brillouin zones are labeled $j=1,2,\ldots,m$, and the principal reduced Brillouin zone (shaded region) has $j=1$.  We show the zone-folded Fermi surface in the principal reduced zone.  Any point ${\bk_a}$ belonging to the $a$th reduced zone can be written $\bk_a = \bk +(a-1) \bq^\ast$. Two such points, $\bk$ and $\bk_2$ are illustrated in the figure.}
\label{fig:bzone}
\end{figure}

When the bond modulation has a period of $m$ unit cells, the Brillouin zone is correspondingly reduced by a factor of $m$ along one direction.    This is illustrated in Fig.~\ref{fig:bzone}.
The mean-field Hamiltonian $H_\mathrm{mf} = H_0 + H_{1\mathrm{bo}}$ can
be written in matrix notation as
\begin{equation}
H_\mathrm{mf} = \sum_{\bk\in \mathrm{BZ}^\prime} \sum_\sigma \Psi_{\bk\sigma}^\dagger {\bf H}_\bk(\bq^\ast) \Psi_{\bk\sigma}
\end{equation}
where $\mathrm{BZ}^\prime$ is the reduced Brillouin zone,  
and $\Psi_\bk$ is a column
vector of length $m$ containing annihilation operators with momenta connected by integer multiples of $\bq^\ast$: 
\begin{equation}
\Psi_{\bk\sigma}  = 
\left [ \begin{array}{c}
c_{\bk_1\sigma} \\
c_{\bk_2\sigma} \\
\vdots \\
c_{\bk_m\sigma}
\end{array}\right  ],
\label{eq:Psidef}
\end{equation}
with
\begin{equation}
\bk_a = \bk + (a-1)\bq^\ast, \quad a \in [1,m].
\end{equation}
In this notation, $\bk_a$ belongs to the $a$th reduced Brillouin zone.
The $m\times m$ matrix ${\bf H}_\bk(\bq^\ast)$ has nonzero elements
\begin{equation}
[{\bf H}_\bk(\bq^\ast) ]_{ab} = \left \{
\begin{array}{ll}
\epsilon_{\bk_a}, & a=b \\
P_{\bk_b}(\bq^\ast),& a=\mod(b,m)+1\\
P_{\bk_a}(\bq^\ast)^\ast,& b=\mod(a,m)+1
\end{array}
\right ..
\label{eq:Hkq}
\end{equation}

Then, the matrix Green's function (including the  impurity self-energy matrix ${\bf \Sigma}_n$) is ${\bf G}(\bk,i\omega_n) = [i\omega_n {\bf 1} - {\bf H}_\bk(\bq^\ast)-{\bf \Sigma}_n]^{-1}$ with
matrix elements
\begin{equation}
G_{ab}(\bk;\omega) = G(\bk_a,\bk_b;\omega).
\end{equation} 
Substituting this into Eq.~(\ref{eq:nematicOP}), the equations for the bond order follow:
\begin{equation}
P^\alpha(\bq^\ast) = -\frac{J_\mathrm{bo} }{mN^\prime} \sum_{\bk\in \mathrm{BZ}^\prime} \sum_{a=1}^m
\eta^\alpha_{\bk_i} G_{a+1 ,a}(\bk;\tau=0^-) 
\label{eq:Pdq}
\end{equation}   
where it is understood that  $a+1 \equiv \mod(a,m) + 1$ and $N^\prime = N/m$ is the number of $\bk$-points in the reduced Brillouin zone.

Without disorder, we can evaluate the Green's function from the eigenvectors and eigenvalues of ${\bf H}_\bk(\bq^\ast)$ to obtain
\begin{equation}
P^\alpha(\bq^\ast) = -\frac{J_\mathrm{bo}}{mN^\prime} \sum_{\bk\in \mathrm{BZ}^\prime} \sum_{a,\ell = 1}^m
\eta^\alpha_{\bk_i} S_{a+1,\ell}(\bk)S_{a,\ell}(\bk) f(E_{\ell \bk}) \\
\end{equation}   
where ${\bf S}(\bk)$ is the matrix of eigenvectors of ${\bf H}_\bk(\bq^\ast)$, and $E_{\ell \bk}$ are the corresponding eigenvalues.  
More generally, once disorder is included, we have 
 \begin{equation}
{\bf G}(\bk;\tau=0^-) = T \sum_n e^{-i\omega_n 0^-} [i\omega_n {\bf 1} - {\bf H}_\bk(\bq^\ast)-{\bf \Sigma}_n]^{-1}.
\label{eq:fullG}
\end{equation}
Substitution of Eq.~(\ref{eq:fullG}) into Eq.~(\ref{eq:Pdq})  generates the self-consistent  equation that must be solved for $P^\alpha(\bq^\ast)$.  The prescription for obtaining ${\bf \Sigma}_n$ within the SCTMA is described in Appendix~\ref{sec:A1}.

\begin{figure}
\includegraphics[width=\columnwidth]{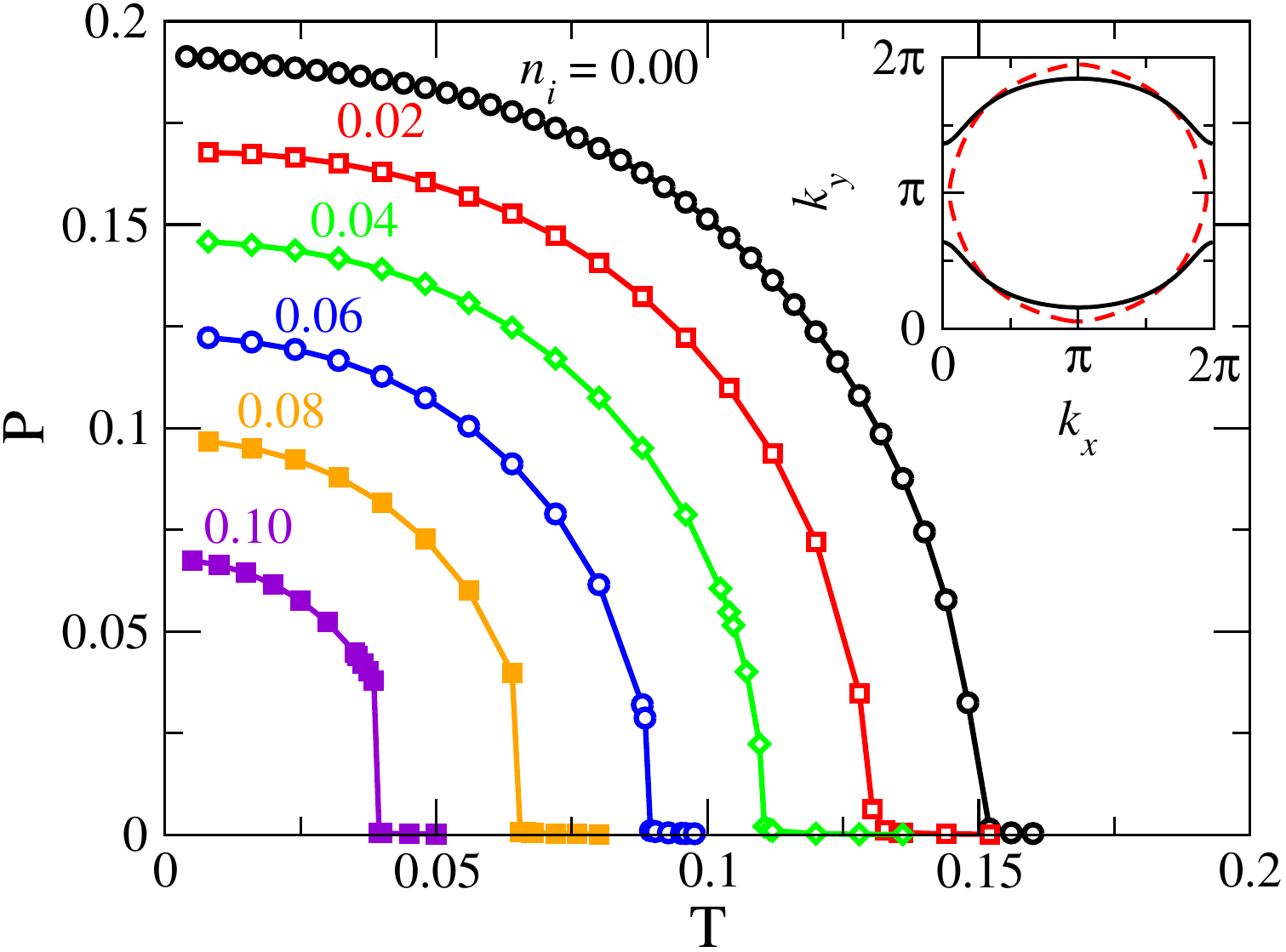}
\caption{Temperature dependence of the  order parameter for the
  $\bq=(0,0)$ nematic instability.  Results are shown for concentrations $n_i$ of strong
  scattering pointlike impurities with potential $V_i = 10.0$.  {\em Inset.}  Comparison of the nematic Fermi surface (red solid) and the bare Fermi surface (black dashed).  Model parameters
  are $t_0 = 1.55$, $t_2= 0.4$, and $J_\mathrm{bo}=1.5$. }
\label{fig:pomeranchuk}
\end{figure}

 As a point of reference, we first revisit the case of nematic order (ie.\ the 
 Pomeranchuk instability) which was
previously studied by Ho and Schofield for a Gaussian distributed
disorder potential.\cite{Ho:2008hn} The Pomeranchuk transition is a
$\bq=0$ instability (so $m=1$), and to obtain it one must tune the Fermi
surface so that it passes near the Brillouin zone boundaries at  $(\pm\pi,0)$ and
$(0,\pm\pi)$.  Here, we take the next-nearest neighbor hopping amplitude $t_2 = 0.4$,
and adjust the filling to obtain the Fermi surface shown by the dashed curve in the inset of Fig.~\ref{fig:pomeranchuk}.
 
 The leading $\bq=0$ instability has a pure $d_{x^2-y^2}$ (or nematic) symmetry, with
\begin{equation}
P_\bk(0) = P[\cos(k_x)-\cos(k_y)],
\end{equation}
where $P \equiv  \sqrt{P^1(\bq^\ast)^2+P^2(\bq^\ast)^2}$. The resulting Fermi surface  in the bond ordered phase is shown by the solid curve in the inset to Fig.~\ref{fig:pomeranchuk}:  the Fermi surface distortion has a clear $d_{x^2-y^2}$ symmetry, with points near $(0,\pm\pi)$ pushed in and points near $(\pm\pi,0)$ pushed away from the Brillouin zone center.

Impurities suppress $P$, and thereby this distortion, as shown in the main panel of Fig.~\ref{fig:pomeranchuk}, and the nematic phase is ultimately destroyed near
$n_i \approx 0.10$.  Ho and Schofield\cite{Ho:2008hn}  noted previously that disorder can change
the order of the transition from second to first in cases where the Fermi surface does not pass
exactly through $(\pm \pi,0)$ and $(0,\pm \pi)$.   This same crossover can be seen in Fig.~\ref{fig:pomeranchuk} at $n_i \approx 0.04$.    In cases where the nematic transition {\em is} second order, $T_\mathrm{bo}$ satisfies the same dependence on the impurity scattering rate $\gamma$ as $d$-wave superconductivity,\cite{Ho:2008hn}  namely Eq.~(\ref{eq:digammaSC}).

\begin{figure}
\includegraphics[width=0.7\columnwidth]{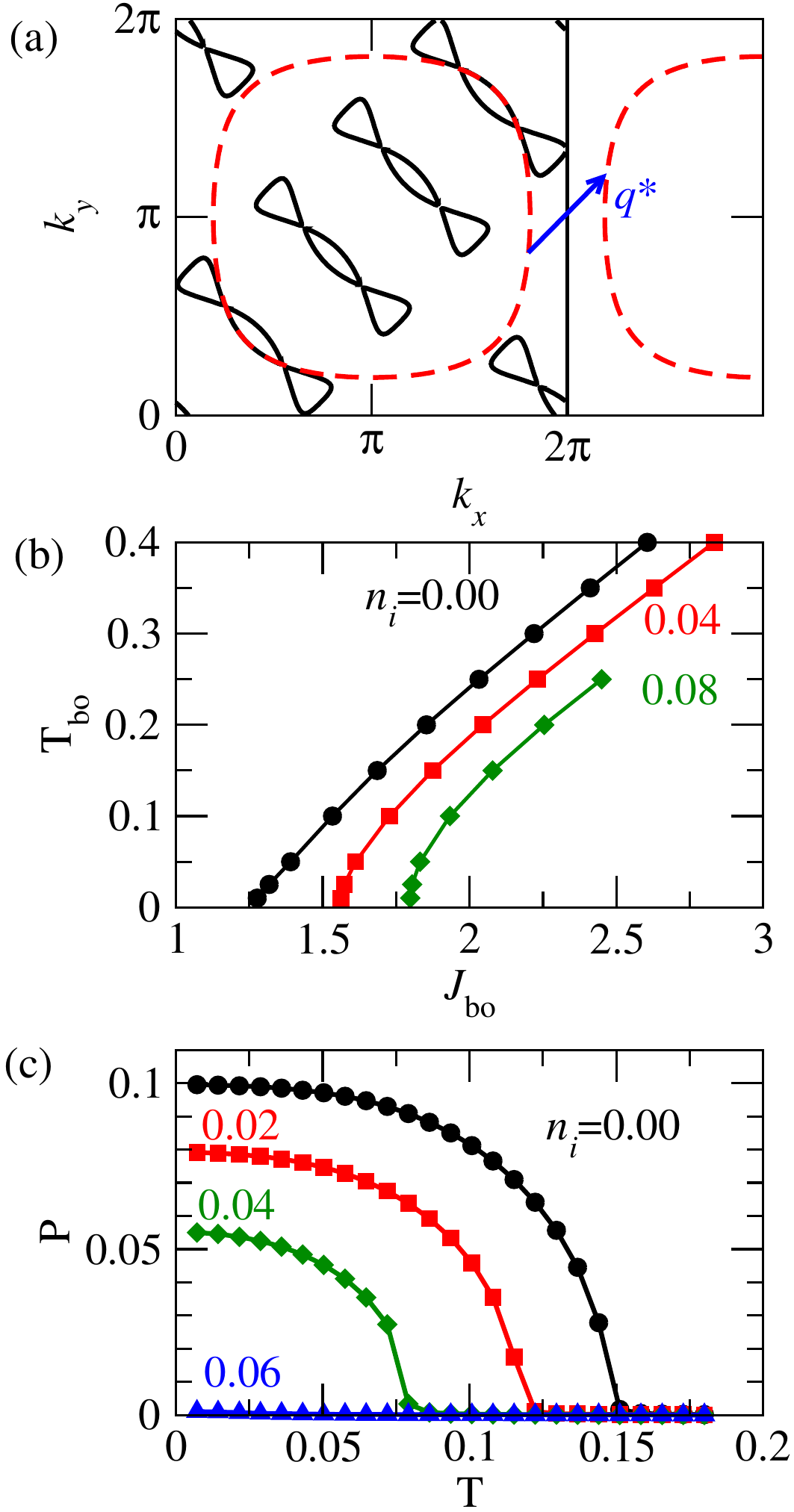}
\caption{Phase diagram for period-5 diagonal  order, with $\bq^\ast = (2\pi/5) (1,1)$.  (a) 
Original (dashed red) and reconstructed (solid black) Fermi surfaces.  The arrow labelled $\bq^\ast$  connects Fermi surface hotspots that are gapped by bond order. (b)
Onset temperature as a function of interaction strength for clean ($n_i=0.00$) and disordered ($n_i=0.04$ and $n_i=0.08$) systems with strong scattering impurities ($V_i=10.0$).
(c) Magnitude of the order parameter as a function of temperature for $J_\mathrm{bo} = 1.68$.    Model parameters are $t_0 = 1.3$ and $t_2 = 0.5$.}
\label{fig:diag_m5}
\end{figure}

Next, we examine the case of diagonal order which, as discussed in the Introduction, was the leading instability in a large number of earlier calculations.  We choose $\bq^\ast = (2\pi/m)(1,1)$ with $m=5$ to give $\bq^\ast$ that is similar in magnitude to what was found earlier.\cite{Holder:2012ks,Bulut:2013,Chowdhury:2014}  To obtain a solution, it is necessary to tune the band parameters so that $\bq^\ast$ connects antiparallel hotspot sections of Fermi surface, as shown in Fig.~\ref{fig:diag_m5}(a).  Near the hotspots, bond order gaps the Fermi surface and thereby reconstructs it as shown in Fig.~\ref{fig:diag_m5}(a).

We remarked earlier that the self-consistent equation for
$P_\bk(\bq)$, Eq.~(\ref{eq:nematic1}), is invariant under $\bk
\rightarrow -\bk-\bq$.  For diagonal order, Eq.~(\ref{eq:nematic1}) is
also invariant under $k_x \leftrightarrow k_y$.  Based on these
two symmetries, we expect  solutions for $P_\bk(\bq^\ast)$ to have
the form
\begin{equation}
P_\bk(\bq^\ast) =  P \left [\cos(k_x +\frac{q_x^\ast}{2}) \pm \cos(k_y+\frac{q_y^\ast}{2}) \right ].
\end{equation}
In our calculations, the solution with the negative sign
is always preferred.  While this solution superficially resembles the $d$-symmetric
order parameter found at $\bq=0$, $P_\bk(\bq^\ast)$ does not have even a
qualitative interpretation as a $d_{x^2-y^2}$ distortion of the Fermi
surface.  Indeed, because of the Brillouin zone folding associated with the finite-$q$ modulation, the reconstructed Fermi surface shown in Fig.~\ref{fig:diag_m5}(a) is quite complicated, with no resemblance to that in Fig.~\ref{fig:pomeranchuk}.

We show the dependence of $T_\mathrm{bo}$
on $J_\mathrm{bo}$ for different impurity concentrations in Fig.~\ref{fig:diag_m5}(b),  
and the $T$-dependence of $P$ for different $n_i$ in
Fig.~\ref{fig:diag_m5}(c).   Similar to the nematic transition, impurities reduce $P$; 
here, however, the nematic transition remains second order as the impurity concentration grows.  As in Sec.~\ref{sec:linearized}, the different $\bq$ components of the order parameter decouple near $T_\mathrm{bo}$, and $T_\mathrm{bo}$ is the same whether the order is uni-directional or bi-directional (checkerboard).  

\begin{figure}
\includegraphics[width=0.8\columnwidth]{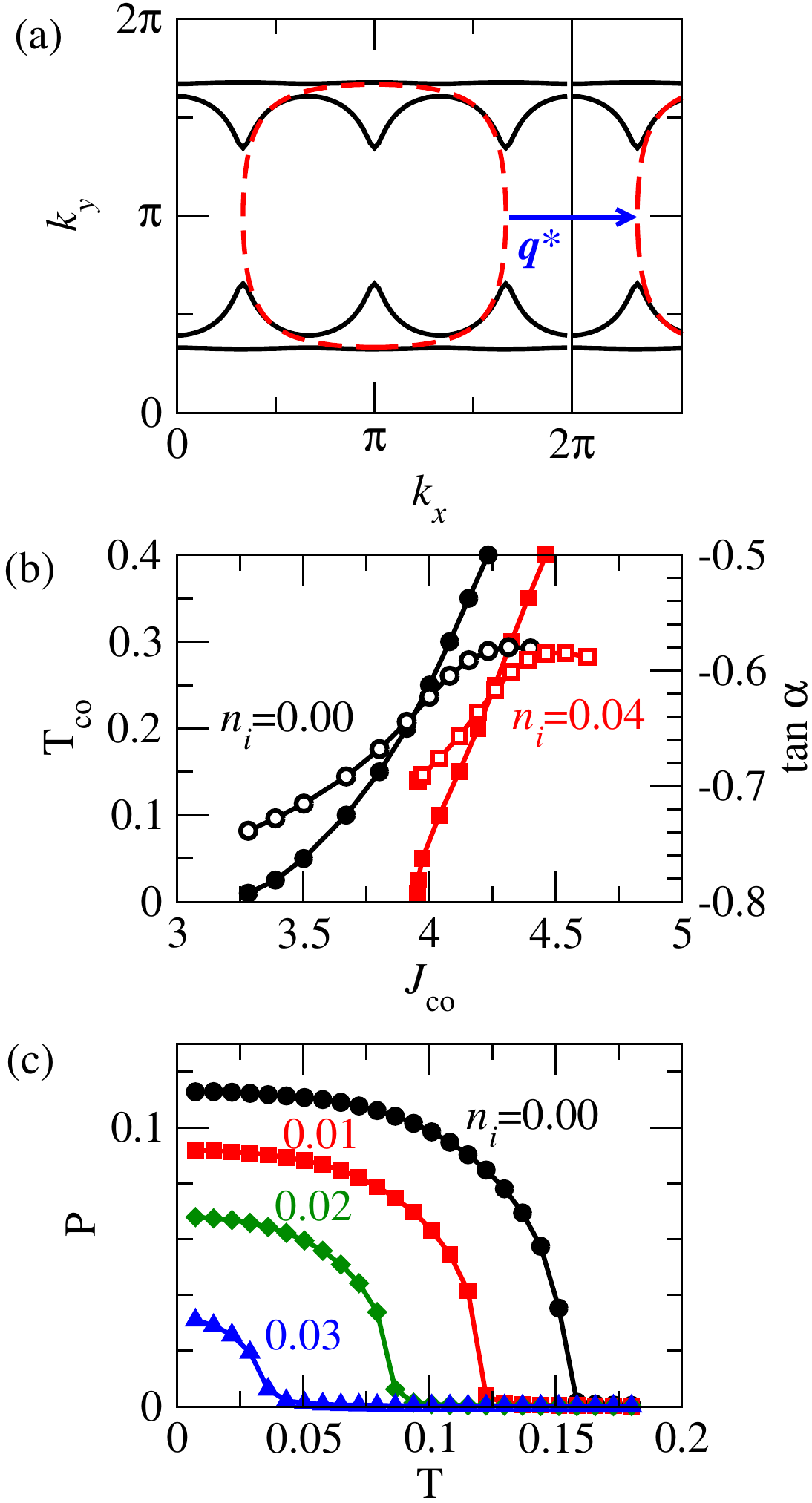}
\caption{Phase diagram for period-3 axial  order, with $\bq^\ast =
  (2\pi/3) (1,0)$.  (a) Original (dashed red) and reconstructed (solid black) Fermi
  surfaces.  The arrow labelled $\bq^\ast$  indicates the bond ordering vector  connecting Fermi surface hotspots.  (b)
  Phase boundaries (solid symbols) for the onset of axial bond order
  for clean ($n_i=0.00$) and disordered ($n_i=0.04$) systems with strong scattering impurities ($V_i=10.0$).  The relative amplitude
  $\tan \alpha$ of  the $k_y$ and $k_x$ components of the order
  parameter, defined by Eq.~(\ref{eq:axial_OP}), is also shown (open symbols).
  (c) Temperature dependence of the
  order parameter $P$ at different impurity concentrations for
  $J_\mathrm{bo} = 3.8$.  The dependence of $\alpha$ on $T$ is very
  weak. Model parameters are $t_0 = 0.4$, $t_2 = 0.7$.}
\label{fig:axial_m3}
\end{figure}

Finally, we consider axial order  with $\bq^\ast = (2\pi/m)(1,0)$, as shown in Fig.~\ref{fig:axial_m3}.  We take $m=3$, which gives $\bq^\ast$ close to that seen experimentally in YBa$_2$Cu$_3$O$_{6+x}$.   Again, it is necessary to tune the band parameters such that  $\bq^\ast$ connects antiparallel portions of the Fermi surface [Fig.~\ref{fig:axial_m3}(a)].  To enhance the susceptibility towards axial order, we have taken $t_2 = 0.7$, which reduces the curvature near the 
Fermi surface hotspots.  (The connection between curvature and $T_\mathrm{bo}$ is discussed, e.g.\ in Ref.~\onlinecite{Metlitski:2010vf}.)  Nonetheless, a rather large $J_\mathrm{bo}=3.8$ is required to obtain a clean-limit transition temperature $T^0_\mathrm{bo}$ that is the same as  in Fig.~\ref{fig:pomeranchuk} for the nematic instability.  

In the axial case, the self-consistent equation for $P_\bk(\bq^\ast)$ is
invariant under $k_x \rightarrow -k_x -q^\ast$ and $k_y\rightarrow
-k_y$.  This implies that  the order parameter  has the
form
\begin{equation}
P_\bk(\bq^\ast) = \sqrt{2} P \left [ \cos(\alpha) \cos(k_x+\frac {q^\ast}{2}) +
  \sin(\alpha) \cos(k_y) \right ],
  \label{eq:axial_OP}
\end{equation}
where $\tan \alpha = P^2(\bq^\ast)/P^1(\bq^\ast)$ is a nonuniversal constant.  

Figure~\ref{fig:axial_m3}(b) shows $T_\mathrm{bo}$ as a function of
$J_\mathrm{bo}$ for two different values of $n_i$, along with $\tan \alpha$.
As before, these results hold for both uni-directional and bi-directional order.
From the plot of $\tan \alpha$, we see that the magnitude of $P^2(\bq^\ast)$ is 60-70\% of the magnitude of $P^1(\bq^\ast)$, that $P^2(\bq^\ast)$ has the opposite sign of $P^1(\bq^\ast)$, and that disorder changes this admixture.  

It is notable that the bond order in the axial and diagonal cases is more rapidly suppressed than in the nematic case, with the axial case the most sensitive to impurities. Equation~(\ref{eq:digamma}) suggests that in the axial and diagonal cases, $T_\mathrm{bo}$ depends on both the Fermi surface curvature and scattering rate.     The nematic transition, on the other hand, approximately  satisfies an equation of the same form as Eq.~(\ref{eq:digammaSC}) for $d$-wave superconductivity,\cite{Ho:2008hn} and at this level of approximation depends only on the scattering rate; nematic order is thus expected to be more robust against impurities than finite-$\bq$ bond order,  consistent with the numerical results shown in Figs.~\ref{fig:pomeranchuk}, \ref{fig:diag_m5}, and \ref{fig:axial_m3}.   Furthermore, comparing the diagonal and axial cases, we note that the Fermi surface curvature in the diagonal case ($\kappa \approx 0.3$) is approximately half that for the  axial case ($\kappa \approx 0.6$), which is consistent with the more rapid suppression of $T_\mathrm{bo}$ in Fig.~\ref{fig:axial_m3} than in Fig.~\ref{fig:diag_m5}.  

Importantly, the scattering rate $\gamma$ also depends  on band structure.    In the strong-scattering limit ($V_i \rightarrow \infty$),
\begin{equation}
\gamma  = \frac{\pi n_i}{N_0},
\end{equation}
where $N_0$ is the density of states at the Fermi energy.   The scattering rate is thus smallest for the nematic order in Fig.~\ref{fig:pomeranchuk} because the Fermi surface passes near van Hove singularities at $(\pm \pi,0)$ and $(0,\pm \pi)$.  We find that for a fixed $n_i$ the scattering rate for the axial case  is roughly twice that for the nematic case, and slightly less than twice that for the diagonal case.    These differences in $\gamma$ are consistent with the different sensitivities to impurities shown in Figs.~\ref{fig:pomeranchuk}, \ref{fig:diag_m5}, and \ref{fig:axial_m3}.   In summary, the sensitivity of bond order to impurities depends on the band structure, both directly through the Fermi surface curvature and indirectly through the scattering rate.   In cuprates, we can thus expect that the sensitivity of charge order to impurities will be doping-dependent.

\subsection{$T_c$ equations for superconductivity in the bond ordered phase}
\label{sec:commensurateSC}
To explore the onset of superconductivity in the bond ordered phase, we consider linearized equations for the pairing instability in the presence of period-$m$ commensurate bond order.
These will give both the superconducting transition temperature $T_ c$, and the $\bk$- and $\bq$-structure of the order parameter $\Delta_\bk(\bq)$ near $T_ c$.  
The mean-field pairing contribution to the Hamiltonian, Eq.~(\ref{eq:Hsc}) is Fourier transformed to obtain
\begin{equation}
H_{1\mathrm{sc}} =   \sum_{\bk,  \bq}  \left[  \Delta_\bk(\bq)^\ast
   c_{-\bk+\bq \downarrow}  c_{\bk \uparrow}  + \mathrm{h.c.}  \right ],
\end{equation}
with $\Delta_\bk(\bq) = \sum_\alpha\eta^\alpha_{\bk}  \Delta^\alpha(\bq)$ and 
\begin{equation}
\Delta^\alpha(\bq) = - J_\mathrm{sc} \frac{1}{N} \sum_{\bk^\prime}
\eta^\alpha_{\bk^\prime} \langle c_{-\bk^\prime+\bq \downarrow}  c_{\bk^\prime \uparrow} \rangle.
\label{eq:SCEq}
\end{equation}
The basis functions $\eta_\bk^\alpha$, defined by Eqs.~(\ref{eq:basis}), are the same as used to describe the bond order.  If the bond order has wavevector $\bq^\ast$, 
 then
the pair order parameter $\Delta^\alpha(\bq)$  must necessarily have Fourier components $\bq = 0,\bq^\ast, 2\bq^\ast, \ldots (m-1)\bq^\ast$.\cite{Zhang:1997,Markiewicz:1998}  Defining 
\begin{equation}
\tilde \Psi_\bk^\dagger = \left [
\begin{array}{cccccccc}
c_{\bk_1\uparrow}^\dagger, & c_{\bk_2\uparrow}^\dagger,& \ldots & c_{\bk_m\uparrow}^\dagger,&
c_{-\bk_1\downarrow}, & c_{-\bk_2\downarrow}, &\ldots & c_{-\bk_m\downarrow}
\end{array} \right  ],
\label{eq:Psitilde}
\end{equation}
where $\bk_a = \bk+(a-1)\bq^\ast$, 
the mean field Hamiltonian containing both superconductivity and bond order is
\begin{equation}
H_\mathrm{mf} = \sum_{\bk\in \mathrm{BZ}^\prime} \tilde \Psi^\dagger_\bk \left [ \begin{array}{cc}
{\bf H}_\bk(\bq^\ast) & {\bf \Delta}_\bk \\
{\bf \Delta}^\dagger_\bk & -{\bf H}^T_{-\bk}(-\bq^\ast)
\end{array}\right ] \tilde \Psi_\bk,
\end{equation}
where ${\bf H}_\bk(\bq^\ast)$ is defined in Eq.~(\ref{eq:Hkq}), and
the $m\times m$ off-diagonal block has matrix elements
\begin{equation}
[{\bf \Delta}_\bk]_{ab} = \sum_\alpha \Delta^\alpha_{a-b} \eta^\alpha_{\bk_a}.
\end{equation}
In this expression, $\Delta^\alpha_{a-b}$ is shorthand for $\Delta^\alpha[(a-b)\bq^\ast]$.
The diagonal elements $[{\bf \Delta}_\bk]_{aa}$ therefore correspond to pairs with zero center-of-mass momentum  belonging to the $a$th reduced Brillouin zone.  We  note that because $m\bq^\ast$ is a reciprocal lattice vector we use terms like $(m-1)\bq^\ast$ and $-\bq^\ast$ interchangeably.

The expectation value in Eq.~(\ref{eq:SCEq}) can be evaluated  to linear order in the pair
amplitude to obtain the eigenvalue equation for the elements of $\mathbf{\Delta}_\bk$ (see Appendix~\ref{sec:B2})
\begin{equation}
\Delta^\alpha_{a-b}  = J_{sc} \sum_{\beta} \sum_{c,d} M_{\alpha (a-b); \beta (c-d)} \Delta^\beta_{c-d}
\label{eq:linearizedTc}
\end{equation}
with 
\begin{eqnarray}
M_{\alpha a; \beta c} &=& -\frac{1}{mN^\prime} \sum_{\bk \in \mathrm{BZ}^\prime} T \sum_n
\sum_{\ell,\ell^\prime=1}^4 \eta^\alpha_{\bk_\ell} \eta^\beta_{\bk_{\ell^\prime}} \nonumber \\
&&\times [i\omega_n - {\bf H}_{\bk}(\bq^\ast)-{\bf \Sigma}_n]^{-1}_{\ell, \ell^\prime} \nonumber \\ 
&&\times [i\omega_n + {\bf H}^T_{-\bk}(-\bq^\ast)+{\bf \Sigma}_n^\ast]^{-1}_{\ell^\prime - c, \ell - a}. 
\label{eq:M}
\end{eqnarray}
In this equation, it is understood that $\ell^\prime - c$ and $\ell - a$ are evaluated modulo $m$.   Furthermore, we have dropped the anomalous impurity self energy ${\bf \tilde \Sigma}_n$: as discussed in Sec.~\ref{sec:linearized}, ${\bf \tilde \Sigma}_n$ vanishes identically in pure $d$-wave superconductors, and as we show below, superconductivity has predominantly $d$-wave symmetry in the bond-ordered phase.  The neglect of ${\bf \tilde \Sigma}_n$ leads us to underestimate $T_c$ slightly; however,  there are two relevant cases where ${\bf \tilde \Sigma}_n=0$ exactly:  (i) $n_i=0$, where the impurity self energy vanishes, and  (ii) cases in which impurities suppress $T_\mathrm{bo}$ such that $T_c > T_\mathrm{bo}$, and the superconductivity is purely $d$-wave.  

\begin{figure}
\includegraphics[width=\columnwidth]{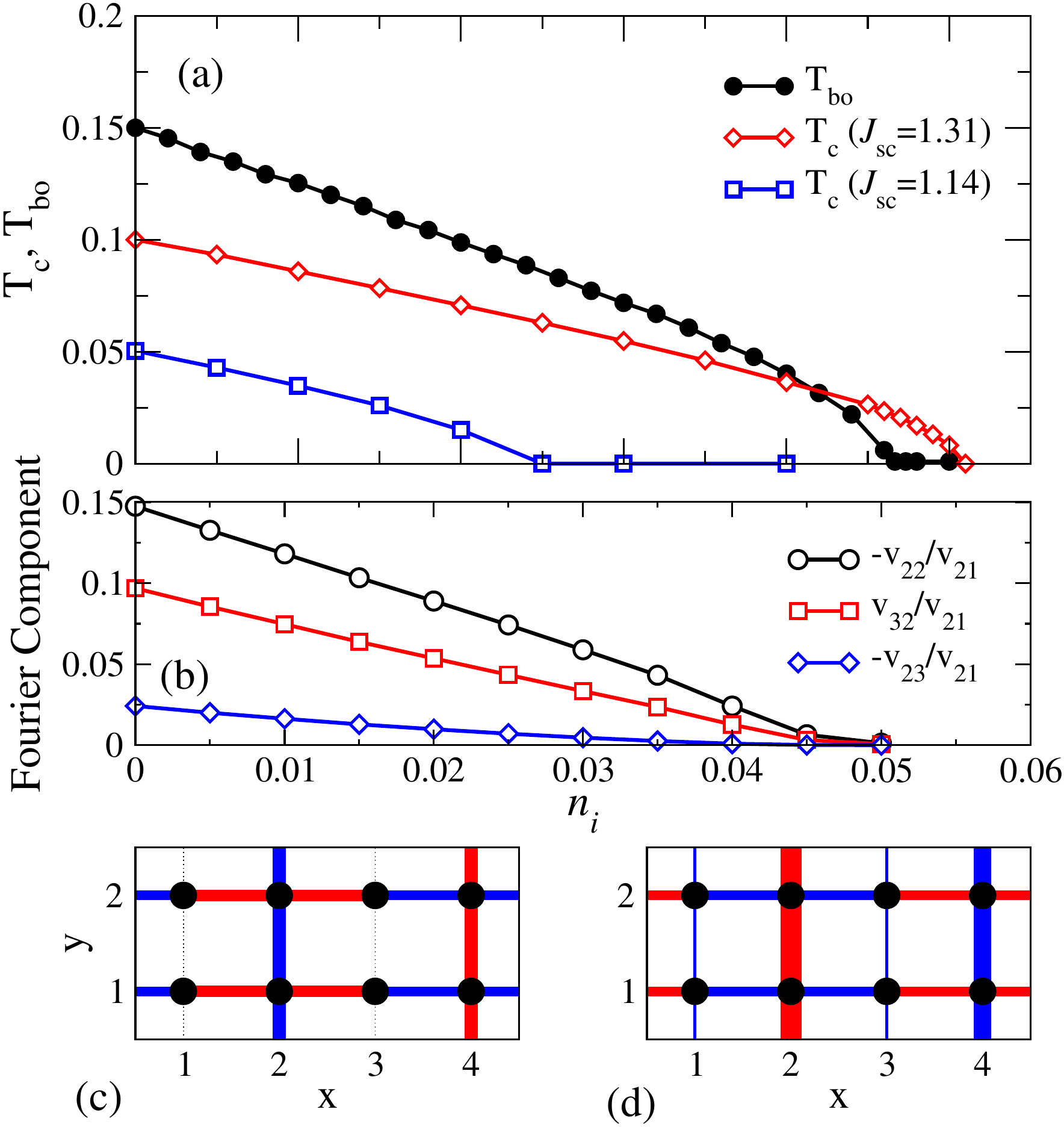}
\caption{Superconducting transition in the bond-ordered phase.   Results are shown for period-4 axial bond order.  (a) Transition temperatures for bond order and superconductivity.  Results are shown for $J_\mathrm{bo} = 2.74$ and two values of the pair interaction $J_\mathrm{sc}$.  Model parameters are otherwise the same as in Fig.~\ref{fig:Tc_vs_Ni}.  (b) Fourier components of the superconducting order parameter at $T_c$, relative to the $d$-wave component $v_0$ [cf.\ Eq.~(\ref{eq:Dk})],  for $J_\mathrm{sc} = 1.31$. (c) Real space pattern of the nonzero-$\bq$ components of $P_{ij}$ at $n_i=0$,  $T=0.10$. The order parameters $P^1(\bq^\ast) = 0.102$ and $P^2(\bq^\ast) = -0.082$ are self-consistently determined.   (d) Nonzero-$\bq$ component of the superconducting order parameter $\Delta_{ij}$ at $n_i=0$ at $T=0.10$.  The pattern is derived from the pairing kernel eigenvector given in Table~\ref{table:eig}.   In (c) and (d), the line thickness and color indicate the magnitude and  sign of the order parameters respectively.  Band parameters are $t_0 =0.85 $ and $t_2=0.5$. }
\label{fig:Tc_vs_Ni2}
\end{figure}

 The kernel $M_{\alpha a;\beta c}$ forms a $16\times 16$ matrix ${\bf M}$ with rows and columns labeled by the composite indices $(\alpha,a)$ and $(\beta,c)$ respectively.
The superconducting instability occurs when the largest eigenvalue of ${\bf M}$ is equal to $1/J_\mathrm{sc}$.  We show the dependence of $T_c$ on impurity concentration in the axial bond-ordered phase in Fig.~\ref{fig:Tc_vs_Ni2}.  The bond order parameter $P_\bk(\bq^\ast)$  is calculated self-consistently, and $T_\mathrm{bo}$ is therefore also shown.  Experimentally, charge order emerges at a higher temperature than superconductivity, although the ratio of $T_ c$ and $T_\mathrm{co}$ is doping dependent and decreases with increasing hole concentration.\cite{Huecker:2014vc}  We therefore show two cases in Fig.~\ref{fig:Tc_vs_Ni2}.   In the first, $T_ c^0 = T_\mathrm{bo}^0/3$, which is comparable to the smallest ratio of $T_ c$ to the charge-ordering temperature seen by x-ray experiments in YBa$_2$Cu$_3$O$_{6+x}$.\cite{Huecker:2014vc}   As $n_i$ increases, $T_\mathrm{bo}$ decreases faster than $T_ c$, although superconductivity is destroyed first.   In the second case, $T_ c^0 = 2T_\mathrm{bo}^0/3$, which is slightly larger than the maximum ratio found in YBa$_2$Cu$_3$O$_{6+x}$.   Here, there is a narrow window over which impurities destroy bond order, but superconductivity remains.  
Note that our calculations explicitly neglect the feedback of superconductivity on bond order, and therefore overestimate $T_\mathrm{bo}$ in regimes where $T_\mathrm{bo} < T_c$ (although $T_c$ {\em is} correctly given).

The eigenvector corresponding to the largest eigenvalue of ${\bf M}$ gives the $\bk$- and $\bq$-space structure of the electron pairs near the superconducting transition.  Given the eigenvector $v_{\alpha a}$, we then have
\begin{equation}
\Delta_\bk(\bq) = \Delta(T)  \sum_{a = 1}^m 
 \delta_{\bq,(a-1)\bq^\ast} \left [ \sum_{\alpha=1}^4 v_{\alpha a}   \eta^\alpha_{\bk} \right ]
\label{eq:Dkq}
\end{equation}
where $\Delta(T)$ is the amplitude of the order parameter and the terms in the square brackets give the $k$-space structure of each Fourier component of $\Delta_\bk(\bq)$.  This equation makes explicit that the pair wavefunction has contributions at multiple center-of-mass momenta.

\begin{table}
\begin{tabular}{l|l|r}
$\alpha$ & $a$ & $v_{\alpha a}$ \\
\hline 
  1   & 1  &  0.489 \\ 
  2   & 1  & -0.701 \\
  3   & 1  & 0.489 \\
  4   & 1  & 0.000 \\
  1   & 2  & 0.000 \\
  2   & 2  & 0.103 \\
  3   & 2  & -0.068 \\
  4   & 2  & 0.000 \\
  1   & 3  & 0.000 \\
  2   & 3  & -0.017 \\
  3   & 3  &  0.000 \\
  4   & 3  & 0.000 \\
  1   & 4  & -0.068 \\
  2   & 4  &  0.103 \\
  3   & 4  & 0.000 \\
  4   & 4  & 0.000 \\
  \hline
\end{tabular}
\caption{Eigenvector $v_{\alpha a}$ corresponding to the largest eigenvalue of the pairing kernel ${\bf M}$ for superconductivity in an axial bond-ordered phase.  Results are for $J_\mathrm{sc} = 1.31$ and $n_i = 0$, and the parameters are otherwise as in Fig.~\ref{fig:Tc_vs_Ni2}.}
\label{table:eig}
\end{table}

To give a concrete example, we consider the order parameter for $n_i = 0$ at temperatures slightly below $T_c^0=0.10$.  We take $m=4$, corresponding to $\bq^\ast = (\frac{\pi}{2},0)$ (see Fig.~\ref{fig:Tc_vs_Ni2} for model parameters). The eigenvector corresponding to the largest eigenvalue of ${\bf M}$  is shown in Table~\ref{table:eig}.  In this case we can simplify Eq.~(\ref{eq:Dkq}) by noting that, within the numerical accuracy of our calculations, $0.701 \approx \sqrt{2} (0.489)$, so that $v_{11}\cos(k_x + \frac \pi 4) + v_{13}\sin(k_x +\frac \pi 4) 
\approx |v_{21}| \cos(k_x)$.  Further, using $\sin(k_x + \frac{\pi}4) = \cos(k_x - \frac{\pi}4)$   we  obtain,
  \begin{eqnarray}
\Delta_\bk(\bq) &=&  \sqrt{2} \Delta(T)  \Big \{  \delta_{\bq,0} |v_{21}| [\cos( k_x) - \cos (k_y) ] \nonumber \\
&&+\delta_{\bq,\bq^\ast} \left [ v_{32} \cos \left (k_x - \frac{\pi}{4}\right ) + v_{22} \cos (k_y )\right ] \nonumber \\
&&+\delta_{\bq,2\bq^\ast} v_{23} \cos (k_y)
\nonumber \\
&&+\delta_{\bq,-\bq^\ast}\left [v_{14} \cos \left ( k_x + \frac {\pi}{4} \right ) + v_{24} \cos (k_y) \right ] 
\Big \}.\nonumber \\
 \label{eq:Dk}
 \end{eqnarray}
 where the factor of $\sqrt{2}$ comes from the definition of $\eta^\alpha_\bk$ and we used the  equivalence of  $-\bq^\ast$ and $(m-1)\bq^\ast$.   All possible harmonics of $\bq^\ast$ are present in $\Delta_\bk(\bq)$; however, the  $\bq=0$ component is largest by far and it has a pure $d_{x^2-y^2}$ symmetry.      

Figure~\ref{fig:Tc_vs_Ni2}(b) shows the  dependence of the different components
of $\Delta_\bk(\bq)$ as a function of impurity concentration for the case $J_\mathrm{sc} = 1.31$.  As bond order is reduced by impurities, the superconducting components at  $\pm \bq^\ast$ and $2\bq^\ast$ make up a progressively smaller fraction of $\Delta_\bk(\bq)$.  When bond order is completely suppressed, these components vanish and the system becomes a dirty $d$-wave superconductor.

The real-space pair amplitudes $\Delta_{ij}$ are more physically transparent than $\Delta_\bk(\bq)$.  Taking nearest neighbor sites $i$ and $j$,
\begin{eqnarray}
\Delta_{ij} &=& \sum_\bq e^{i\bq \cdot \bR} \frac{1}{N} \sum_{\bk} e^{i(\bk-\bq/2)\cdot \br_{ij}}  \Delta_\bk(\bq), 
\end{eqnarray}
where $\bR =  (X,Y) =  (\br_i + \br_j)/2 $ and $\br_{ij} = \br_i -\br_j$.
We obtain
\begin{eqnarray}
\Delta_{j+x\,j} &=& \sqrt{2}\Delta(T) \Bigg [ \frac{1}{2}v_{21} + v_{32} \cos \left( \frac{\pi X}{2} \right ) \Bigg ] \\
\Delta_{j+y\,j} &=& \sqrt{2}\Delta(T)  \Bigg [ - \frac{1}{2}v_{21} +v_{22} \cos \left( \frac{\pi X}{2} \right )
\nonumber \\ &&+ \frac{1}{2} v_{23} \cos(\pi X) \Bigg ].
\end{eqnarray}
Similarly, the real-space bond order parameter $P_{ij}$ can be obtained by inverting Eq.~(\ref{eq:nematic0}).  Plots of $P_{ij}$ and $\Delta_{ij}$ with the homogeneous $\bq=0$ component removed are given in Figs.~\ref{fig:Tc_vs_Ni2}(c)  and (d) respectively.  These figures explicitly show that the spatial modulations of the pairing amplitude  and bond order are correlated.

\section{Discussion}
\label{sec:discuss}
The results in Figs.~\ref{fig:Tc_vs_Ni} and \ref{fig:Tc_vs_Ni2} suggest a way to probe possible relationships between charge order and the pseudogap, namely to track the dependence of the pseudogap on zinc doping.   We can compare to a number of early experiments that explored exactly this, principally in YBa$_2$Cu$_4$O$_8$, which is often seen as a model underdoped cuprate because it is stoichiometric.
We note, however, a well-known and persistent problem that because the pseudogap appears as a crossover rather than a phase transition, the identification of the relevant temperature scale(s) depends on the experimental technique, and on how the temperature scales are defined.

 Julien {\em et al.}\cite{Julien:1996,Timusk:1999wp} noted that early experiments on underdoped YBa$_2$Cu$_3$O$_{6+x}$ and on YBa$_2$Cu$_4$O$_8$ found two distinct temperature scales, with dramatically different responses to Zn impurities.   The higher scale,  $T^\ast \sim 200$-300 K, was seen originally in Knight shift measurements\cite{Alloul:1989} that indicated a reduction of available spin excitations below $T^\ast$. Later optical conductivity measurements showed that there is an accompanying reduction in available charge excitations.\cite{Homes:1993,Timusk:1999wp}  Experimentally, $T^\ast$ was found to be independent of Zn concentration.\cite{Alloul:1991,Zheng:1993,AlloulRMP:2009}  The lower temperature scale $T^\prime \sim 150$ K was observed as a downturn in the NMR relaxation rate\cite{Warren:1989,Zheng:1996wh} $1/T_1T$ and in the in-plane Hall coefficient.\cite{Mizuhashi:1995} The downturn in the Hall coefficient has recently been tied to the onset of charge order at $T_\mathrm{co}$.\cite{LeBoeuf:2011,Chang:2012vf}  This lower temperature is rapidly suppressed by Zn doping.\cite{Zheng:1993,Zheng:1996wh}

In particular, Zn doping experiments\cite{Miyatake:1991,Zheng:1993,Zheng:1996wh} on YBa$_2$(Cu$_{1-z}$Zn$_z$)$_4$O$_8$ found that $T_c$ was  suppressed from $T_c^0 \approx 80$ K for $z=0$ to $T_c = 0$ K for $z \sim 0.03$, while $T^\prime$ was suppressed much faster, \cite{Zheng:1996wh}  from 150 K at $z=0$ to 0 K at $z \approx 0.02$.   Similar results\cite{Zheng:1993,Zheng:1996wh} were found for YBa$_2$(Cu$_{1-z}$Zn$_z$)$_3$O$_{6.63}$. 
Qualitatively, these are consistent with the suppression of $T_\mathrm{bo}$ shown in Fig.~\ref{fig:Tc_vs_Ni2}.  To make a quantitative comparison, we note that Zn substitutes preferentially for Cu sites in the CuO$_2$ planes so that  in YBa$_2$(Cu$_{1-z}$Zn$_z$)$_4$O$_8$ the Zn concentration per planar Cu is  $2z$.   With this in mind, it is clear that our calculations overestimate reduction of superconductivity by disorder, relative to experiments.  This is a known problem with disorder-averaged calculations of $T_c$ in cuprates, which neglect spatial inhomogeneity of the order parameter.\cite{AlloulRMP:2009}

Although we have suggested that $T^\prime$ and $T_\mathrm{bo}$ may be the same temperature scale, we emphasize that a direct comparison between the suppression of $T^\prime$ by Zn in cuprates and the suppression of $T_\mathrm{bo}$ by impurities in our calculations is not straightforward.  In particular, Zn impurities are known to nucleate magnetic moments locally around each impurity site.\cite{Alloul:1991}    NMR $T_1$ measurements are certainly affected by these moments, and indeed it has been suggested that they are sufficient to explain the doping dependence of $1/T_1T$.\cite{AlloulRMP:2009}   
 In practice, it may be difficult to disentangle the contributions of local moments and impurity scattering to the suppression of charge order in the cuprates.

Finally, we remark that the rapid suppression of charge order in YBCO by Zn impurities is in contrast to the apparent enhancement of stripe correlations\cite{Vojta:2009,Schmid:2013} in Zn-doped La$_{2-x}$Sr$_x$CuO$_4$ (this point was also made in Ref.~\onlinecite{Huecker:2014vc}).   We take this as further evidence that the physics underlying charge order in YBCO and BSCCO is different than that in the La-based cuprates.

\section{Conclusions}
We have studied the effects of strong-scattering pointlike impurities on charge order and superconductivity in the cuprate superconductors.  Calculations were based on a one-band model in which bond order is the analogue of charge order in the cuprates.  Impurity effects were described  with a self-consistent t-matrix approximation.

Our main observation is that $d$-wave superconductivity is more robust against  impurities than bond order;   this implies that charge order in the cuprates should be more rapidly reduced by Zn substitution than supercondutivity, even though the onset temperature for charge order is higher than $T_c$.  Interestingly, the sensitivity of bond order to impurities is not directly connected to the symmetry of the order parameter, but occurs because charge order arises from only small ``hotspot" regions of the Fermi surface.

Experimentally, both the pseudogap and stripe phase in cuprate high temperature superconductors are insensitive to Zn doping.  This is inconsistent with simple scenarios in which charge order contributes directly to the pseudogap.

\section*{Acknowledgments}
We thank M.-H.\ Julien and D.\ G.\ Hawthorn for helpful conversations.
W.A.A. acknowledges support by the Natural Sciences and Engineering Research Council (NSERC)
of Canada.  A.P.K. acknowledges support by the Deutsche Forschungsgemeinschaft through TRR 80.

\appendix

\section{Impurities in the bond-ordered phase}

\label{sec:A}

We use the self-consistent t-matrix approximation (SCTMA) to obtain an expression for the self energy ${\bf \Sigma}(i\omega_n)$ due to the impurities.  The SCTMA gives the disorder-averaged Green's function and is exact in the limit where the impurity concentration $n_i$ is small.\cite{BalatskyRMP:2006}  Apart from the complications arising from the charge order, our approach is standard. 

The derivations in this appendix have three parts.  In Appendix~\ref{sec:A2}, the scattering self energy for weak bond order is obtained to linear order in $P_\bk(\bq)$; this is used to obtain the self-consistent equations for $T_\mathrm{bo}$.   These are solved in Appendix~\ref{sec:A3} to find an approximate analytic expression for $T_\mathrm{bo}$. Finally,  in Appendix~\ref{sec:A1} we find the self energy for the case of arbitrarily strong bond order with period-$m$ commensurability.  In this case, ${\bf \Sigma}(\omega)$ is an $m\times m$ matrix.  

\subsection{Linearized Results near $T_\mathrm{bo}$}
\label{sec:A2}
In this section, we derive Eq.~(\ref{eq:Fab}), along with expressions for the self-energy components $\Sigma_n$ and $\Sigma_n^\bq$ which are valid to zeroth and first order in $P_\bk(\bq)$ respectively.
We consider a dilute distribution of $N_i$ pointlike impurities.  Each impurity is assumed to shift the potential on a lattice site by $V_i$, and we will make use of the assumption that $n_i \equiv N_i/N \ll 1$, where $N$ is the number of lattice sites.   The  potential energy of electrons interacting with the  impurities is
\begin{equation}
\hat V = V_i \sum_{I=1}^{N_i}  \hat n_{\bR_I} = V_i  \sum_{\bk, \bk^\prime} \frac{1}{N} \sum_{I=1}^{N_i} e^{-i (\bk-\bk^\prime)\cdot \bR_I} \sum_\sigma c^\dagger_{\bk \sigma} c_{\bk^\prime \sigma}
\label{eq:V0}
\end{equation}
where $\bR_I$ is the position of impurity $I$ and $\hat n_{\bR_I}$ is the electron 
charge density operator on 
site $\bR_I$.  

\begin{figure}
\includegraphics[width=\columnwidth]{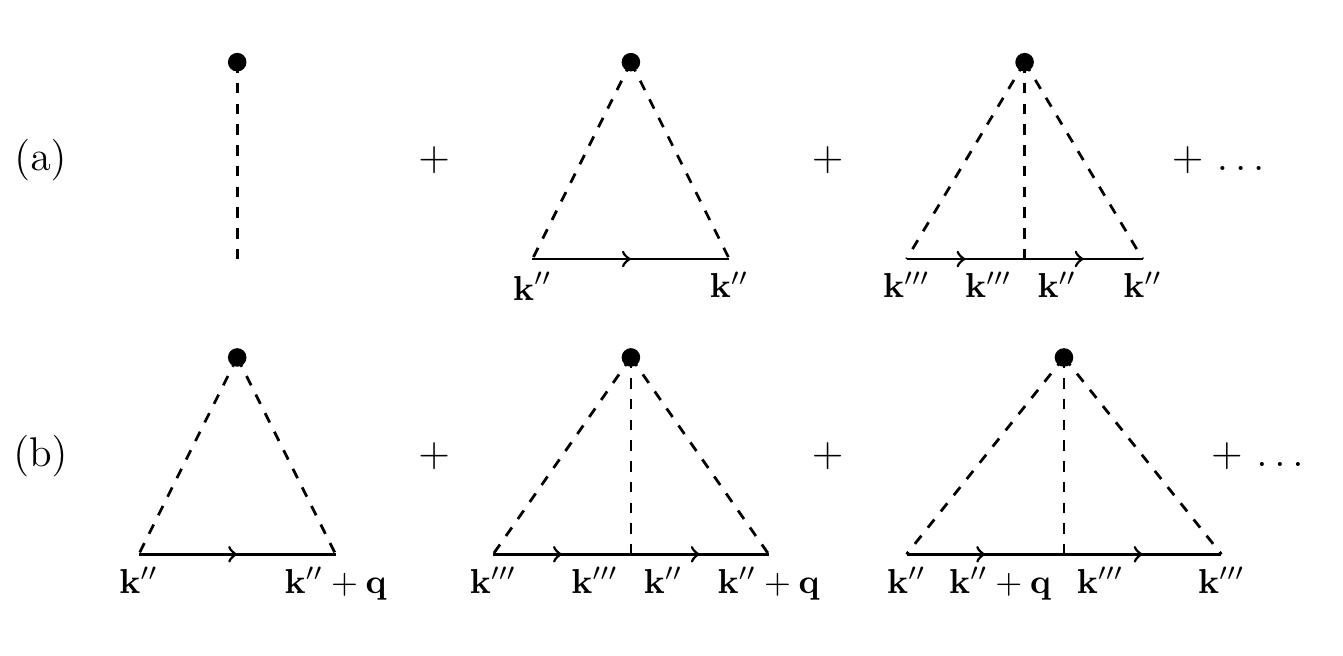}
\caption{SCTMA  diagrams contributing to the self energy to (a) zeroth order and (b) first order in $P_\bk(\bq)$.  Internal wavevectors $\bk^{\dprime}$ and $\bk^{\prime\prime\prime}$ are summed over.  Dashed lines  represent scattering by the impurity potential $V_i$, and solid lines represent the Green's functions $G(\bk^\dprime,\bk^\dprime;i\omega_n)$, $G(\bk^\dprime+\bq,\bk^\dprime;i\omega_n)$, etc., as indicated by the momentum labels.
}
\label{fig:linearizedSE}
\end{figure}

The impurity self energy is obtained by disorder-averaging over the possible positions $\bR_I$ of each impurity, and retaining all irreducible diagrams that are first order in $n_i$.   Figure~\ref{fig:linearizedSE} shows diagrammatic contributions to  $\Sigma_n$ and $\Sigma^\bq_n$.  The first term in Fig.~\ref{fig:linearizedSE}(a) is
\begin{equation}
   \frac{V_i}{N} \sum_{I=1}^{N_i} \left \langle  e^{-i (\bk-\bk^\prime)\cdot \bR_I} \right \rangle  = n_i V_i \delta_{\bk,\bk^\prime}, 
\end{equation}
where $\langle \ldots \rangle = \frac{1}{N} \sum_{\bR_I} [\ldots]$  is the average over all possible positions for the $I$th impurity.  To obtain the second  term in Fig.~\ref{fig:linearizedSE}(a), we keep only second-order scattering contributions in which both impurity lines are from the same impurity.   This gives
\begin{eqnarray}
&& \frac{ V_i^2 }{N^2} \sum_{I,J=1}^{N_i} \delta_{I,J} \sum_{\bk^\dprime} G(\bk^\dprime,\bk^\dprime; i\omega_n)
  \nonumber \\ &&\times 
  \left \langle  e^{-i (\bk-\bk^\dprime)\cdot \bR_I} e^{-i (\bk^\dprime-\bk^\prime)\cdot \bR_J} \right \rangle  \nonumber \\
 &=& n_i V_i^2 \delta_{\bk,\bk^\prime}{\cal G}^0_n
 \end{eqnarray}
where
$ {\cal G}^0_n = N^{-1}\sum_{\bk^\dprime} G(\bk^\dprime,\bk^\dprime; i\omega_n)$.  Following this procedure, the $j$th order diagram is then $ n_i [V_i {\cal G}^0_n]^j\delta_{\bk,\bk^\prime}$, and
the sum of  diagrams to infinite order is $\Sigma_n \delta_{\bk,\bk^\prime}$, where
\begin{equation}
\Sigma_n  = \frac{n_i V_i}{1-V_i{\cal G}^0_n}.
\label{eq:Sigma0}  
\end{equation}

The sum of diagrams of the type shown in Fig.~\ref{fig:linearizedSE}(b) can be obtained in similar fashion.  There are $j-1$ terms at $j$th order in $V_i$:  each of these terms contains $j-2$ factors of ${\cal G}^0_n$ and one factor of $ {\cal G}^\bq_n= N^{-1}\sum_{\bk^\dprime} G(\bk^\dprime+\bq,\bk^\dprime; i\omega_n)$.  The sum of diagrams is $\Sigma^\bq_n \delta_{\bk^\prime,\bk+\bq}$ where
\begin{eqnarray}
\Sigma_n^\bq &=& n_i V_i^2 {\cal G}^\bq_n \left [ 1 + 2 V_i {\cal G}^0_n
+ 3 V_i^2 {{\cal G}^0_n}^2
+\ldots \right ] \nonumber \\
&=&  \frac{n_i V_i^2 {\cal G}^\bq_n }{[1-V_i{\cal G}^0_n]^2}.
\label{eq:SigmaQ}
\end{eqnarray}

The equations for $\Sigma_n$ and $\Sigma^\bq_n$ are made self-consistent by obtaining equations for ${\cal G}^0_n$ and  ${\cal G}^\bq_n$.
These come from the  equations of motion for the Green's function,
\begin{widetext}
\begin{eqnarray}
\left [ \begin{array}{cc}
G(\bk,\bk; i\omega_n) & G(\bk,\bk+\bq; i\omega_n) \\
G(\bk+\bq,\bk; i\omega_n) & G(\bk+\bq,\bk+\bq; i\omega_n)
\end{array}\right ] 
\left [ \begin{array}{cc}
i\omega_n - \epsilon_\bk -\Sigma_n & -P_\bk(\bq)-\Sigma^{\bq}_n \\
-P_\bk(\bq) -\Sigma^{\bq}_n & i\omega_n - \epsilon_{\bk+\bq} -\Sigma_n 
\end{array}\right ]  
=
\left [ \begin{array}{cc}
1 & 0 \\ 0 & 1
\end{array}\right ], 
\end{eqnarray}
\end{widetext}
from which,
\begin{eqnarray}
G(\bk,\bk;i\omega_n) &=& \frac{1}{i\omega_n - \epsilon_\bk - \Sigma_n} \label{eq:zeroG} \\
G(\bk+\bq,\bk; i\omega_n) 
&=& \frac{P_\bk(\bq) + \Sigma^\bq_n}{(i\omega_n-\epsilon_{\bk+\bq}-\Sigma_n)
(i\omega_n-\epsilon_{\bk} -\Sigma_n)},\nonumber \\
\label{eq:linearG}
\end{eqnarray}
 to linear order in $P_\bk(\bq)$.

 Equations (\ref{eq:Sigma0}) and  (\ref{eq:zeroG}) form a closed set of self-consistent equations for $\Sigma_n$.   Once $\Sigma_n$ is known, Eq.~(\ref{eq:SigmaQ}) and  Eq.~(\ref{eq:linearG})  can then be solved self-consistently for $\Sigma^\bq_n$:
 \begin{equation}
 \Sigma^\bq_n =  \frac{n_iV_i^2  \tilde {\cal G}^\bq_n }
 {\left [ 1-V_i{\cal G}^0_n \right ]^2 - n_i V_i^2 {\cal G}^\prime_n }
 \label{eq:SigmaQ2}
 \end{equation}
 with 
 \begin{equation}
 {\cal G}^\prime_n = \frac{1}{N} \sum_\bk \frac{1}{(i\omega_n - \epsilon_\bk -\Sigma_n)(i\omega_n - \epsilon_{\bk+\bq} -\Sigma_n)},
\end{equation}
and 
\begin{equation}
 \tilde {\cal G}^\bq_n = \frac{1}{N} \sum_\bk \frac{P_\bk(\bq)}{(i\omega_n - \epsilon_\bk -\Sigma_n)(i\omega_n - \epsilon_{\bk+\bq} -\Sigma_n)}.
\end{equation}
From Eq.~(\ref{eq:SigmaQ2}), one can express $\Sigma^\bq_n$ as 
\begin{equation}
\Sigma^\bq_n = \sum_\alpha {\cal S}^{\bq,\alpha}_n P^\alpha(\bq),
\label{eq:Sigmaq}
\end{equation}
where $P^\alpha(\bq)$ is defined by Eq.~(\ref{eq:nematicOP}).
Once $\Sigma_n$ and $\Sigma_n^\bq$ are known, we substitute Eq.~(\ref{eq:linearG}) into  Eq.~(\ref{eq:nematicOP}) to obtain
\begin{equation}
F_{\alpha\beta}(\bq) = -\frac{T}{N} 
\sum_{\bk,n}   
 \frac{ \eta^\alpha_{\bk} (\eta^\beta_{\bk} + {\cal S}^{\bq,\beta}_n) }
 {(i\omega_n - \epsilon_{\bk+\bq} -  \Sigma_n)
 (i\omega_n - \epsilon_{\bk} -  \Sigma_n)},
\end{equation}
which is  Eq.~(\ref{eq:Fab}) in the text.
In the clean limit ($\Sigma_n = \Sigma^\bq_n =  0$), this reduces to 
\begin{equation}
F_{\alpha\beta}(\bq) = -\frac{1}{N}
\sum_{\bk} \eta^\alpha_{\bk} \eta^\beta_{\bk}
 \frac{f(\epsilon_{\bk+\bq})-f(\epsilon_{\bk})}{\epsilon_{\bk+\bq} - \epsilon_{\bk}}.
 \label{eq:Fab0}
\end{equation}

\subsection{Analytic approximation for $T_\mathrm{bo}$}
\label{sec:A3}

\begin{figure}
\includegraphics[width=\columnwidth]{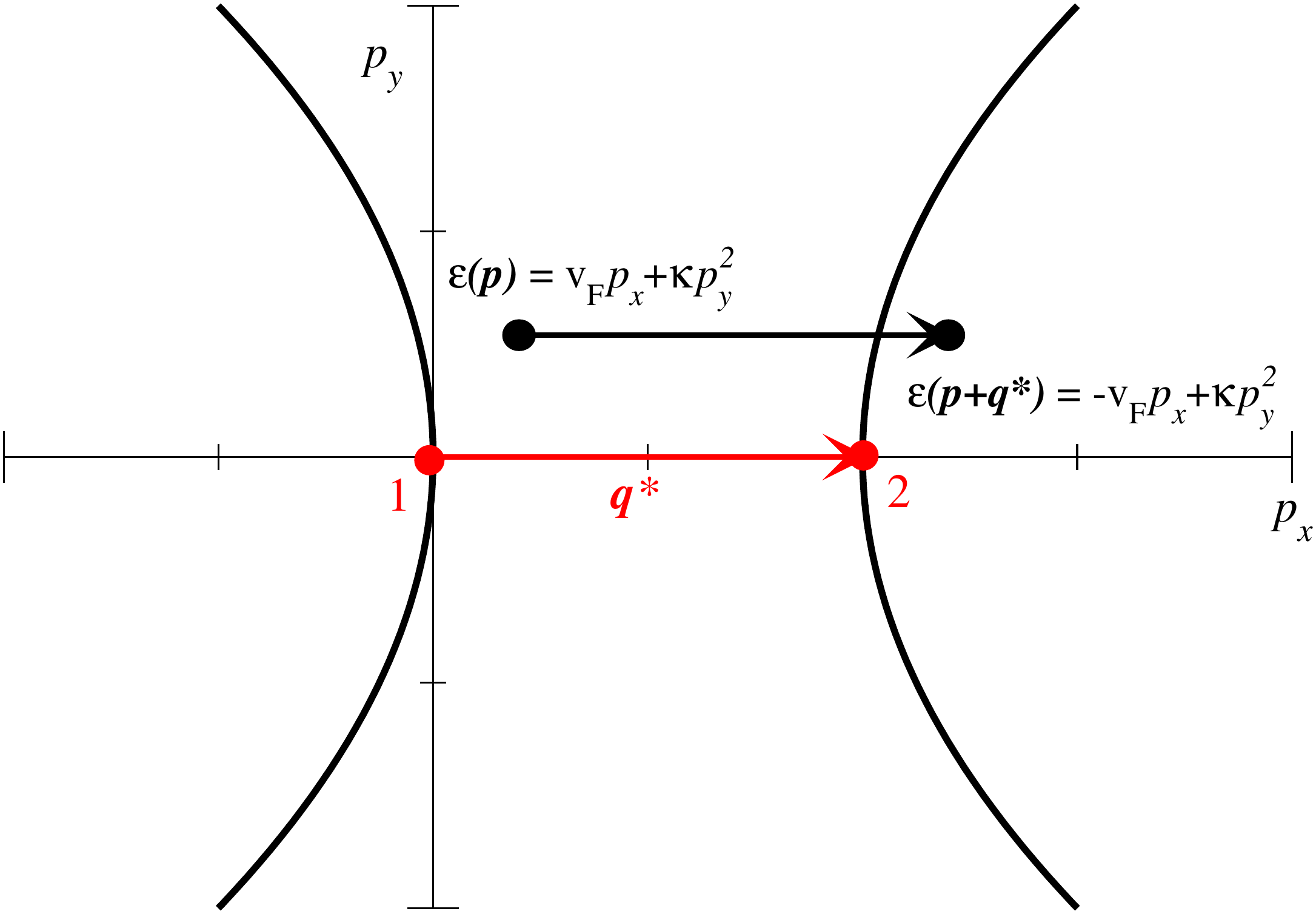}
\caption{Illustration of the dispersion $\epsilon(\bk)$ near the Fermi surface points nested by the bond ordering wavevector $\bq$.  Solid black curves are Fermi surface segments and $\bq^\ast$ connects hotspot points 1 and 2, at which the joint density of states is large.  The dispersions in the neighborhood of points 1 and 2 are $\epsilon(\bq) = v_F p_x + \kappa p_y^2$ and
 $\epsilon(\bq+\bq^\ast) = - v_F p_x + \kappa p_y^2$ respectively.  In this figure, the origin of
 the coordinate system is hotspot 1, not the center of the Brillouin zone.}
\label{fig:analytic}
\end{figure}

We begin with Eq.~(\ref{eq:Fab}) for the bond ordering kernel $F_{\alpha\beta}(\bq^\ast)$ and
make a number of simplifications. First,  we assume that $\bq^\ast$ nests two Fermi surface hotspots, labelled 1 and 2 in  Fig.~\ref{fig:analytic}, that are characterized by anti-parallel Fermi velocities $v_F$ and by curvatures $\kappa$.   By expanding the dispersion around the hotspots we obtain
\begin{equation}
\epsilon_\bk = v_F p_x + \kappa p_y^2;\quad  \epsilon_{\bk+\bq^\ast} = - v_F p_x + \kappa p_y^2,
\end{equation}
where $\bq$ is the wavevector measured relative to hotspot 1 (whereas $\bk$ is relative to the Brillouin zone center).

Then, we make the approximation that the scattering self energy is piecewise constant, so $\Sigma_n = \Delta\mu - i\gamma \mbox{sgn}(\omega_n)$ and absorb the real part $\Delta\mu$ into the chemical potential.  This approximation is not entirely justified, owing to a nearby van Hove singularity in the density of states; however, we have found that adding a weak linear energy dependence to $\gamma$ does not change our answers appreciably.

For definiteness, we will consider period-$m$ axial order, with 
$q_x = q^\ast= 2\pi/m$ and $q_y = 0$.  We know from numerics that only the basis functions
\begin{equation}
\eta^1_\bk = \sqrt{2}\cos(k_x+\frac {q^\ast}2);\quad \eta^2_\bk = \sqrt{2}\cos(k_y)
\end{equation}
contribute to $P_\bk(\bq^\ast)$, so we restrict our discussion to the $2\times 2$ subspace in which 
$\alpha,\beta \in \{1,2\}$.
For axial order, hotspot 1 is, in the original coordinate system, at $k_x = 2\pi-q^\ast/2$ and $k_y=\pi$ (see, for example, Fig.~\ref{fig:axial_m3}(a)), and we approximate $\eta^1_{\bk}$ and $\eta^2_{\bk}$ by their values at this point.  We thus obtain
\begin{equation}
\eta^1_{\bk} \approx \sqrt{2} ;\quad \eta^2_{\bk} \approx -\sqrt{2}.
\label{eq:etahotspot}
\end{equation}

Numerically, we find that $\Sigma^{\bq^\ast}_n$ is small and, neglecting it, we obtain
\begin{equation}
F_{11}(\bq^\ast) = F_{22}(\bq^\ast) = -F_{21}(\bq^\ast) = -F_{12}(\bq^\ast).
\label{eq:Fapprox}
\end{equation}
with 
\begin{equation}
F_{11}(\bq^\ast) \approx -\frac{2T}{N}  
\sum_{\bq,n}   
 \frac{1 }{(i\tilde \omega_n - \epsilon_{\bq+\bq^\ast} )
 (i\tilde \omega_n - \epsilon_{\bq})},
  \label{eq:Fab2}
\end{equation}
and $\tilde \omega_n = \omega_n + \gamma \mbox{sgn}(\omega_n)$.

According to Eq.~(\ref{eq:linearized}), the onset of bond order occurs when the largest eigenvalue
of ${\bf F}(\bq^\ast)$ is equal to $1/J_\mathrm{bo}$.  The eigenvectors of ${\bf F}(\bq^\ast)$ are 0 and $2F_{11}(\bq^\ast)$, so $T_\mathrm{bo}$ satisfies
\begin{equation}
1 = 2J_\mathrm{bo}F_{11}(\bq^\ast).
\label{eq:TcEq}
\end{equation}
The corresponding eigenvector of ${\bf F}(\bq^\ast)$ is $(1,-1)^T$, which gives the $d_{x^2-y^2}$-like
solution $P^1(\bq^\ast) = -P^2(\bq^\ast)$, or
\begin{equation}
P_\bk(\bq^\ast) = P^1(\bq^\ast) ( \eta^1_\bk - \eta^2_\bk),
\end{equation}
 similar to that found numerically.   Our goal is now to estimate $F_{11}(\bq^\ast)$.

Transforming the summation over $\bq$ to an integral, Eq.~(\ref{eq:Fab2})  becomes
\begin{eqnarray}
F_{11}(\bq^\ast) &=& -2T\int_{0}^{\pi}  \frac{d p_y}{\pi} \int_0^{\pi v_F} \frac{dx}{v_F\pi} \sum_{n=-\infty}^{\infty}   \nonumber \\
&& \times  
\frac{1}{(x + i\tilde \omega_n - \kappa p_y^2)(x-i\tilde \omega_n + \kappa p_y^2)} 
\end{eqnarray}
where $x = v_F p_x$.  The term $\pi v_F$ is a large-energy cutoff, and is assumed much bigger than any other energy scale in the calculation.

Evaluating the integral over $x$, and substituting into Eq.~(\ref{eq:TcEq})  gives an equation
for the bond ordering temperature,
\begin{equation}
1 = 4J_\mathrm{bo} \frac{T_\mathrm{bo} }{v_F\pi} \int_{0}^{\pi} \frac{d p_y}{\pi}  \sum_{n=0}^{\infty}  
\frac{\tilde \omega_n}{\tilde \omega_n^2 + \kappa^2p_y^4} \tan^{-1} \left ( \frac{\pi v_F}{\tilde \omega_n} \right )
\end{equation}
where we have dropped a small logarithmic correction that vanishes in the limit $\pi v_F \gg \kappa p_y^2$.

A similar equation holds for the clean limit transition temperature $T_\mathrm{bo}^0$ provided we replace $\tilde \omega_n$ by $\omega_n$.  Setting these two equations equal to each other, we 
obtain
\begin{eqnarray}
\sum_{n=0}^{\Lambda_c} \int_0^\pi \frac{dp_y}{\pi} \frac{n + \frac 12 +\tilde \gamma}{ (n + \frac 12 +\tilde \gamma)^2 + \tilde \kappa_c p_y^2} \nonumber \\
=
\sum_{n=0}^{\Lambda_0} \int_0^\pi \frac{dp_y}{\pi} \frac{n + \frac 12 }{ (n + \frac 12 )^2 + \tilde \kappa_0 p_y^2}
\end{eqnarray}
where $\tilde \kappa_c = \kappa/(2\pi T_\mathrm{bo})$ and $\tilde \kappa_0= \kappa/(2\pi T_\mathrm{bo}^0)$, and
$\Lambda_c  = v_F/2T_\mathrm{bo}$ and $\Lambda_0 = v_F/2T_\mathrm{bo}^0$.  The 
cutoffs $\Lambda_c$ and $\Lambda_0$ come from approximating $\tan^{-1} (x) \sim (\pi/2) \Theta(x-1)$, with $\Theta(x)$ the Heavyside step function.  Performing the sum over Matsubara frequencies, we obtain the final result, Eq.~(\ref{eq:digamma}).

\subsection{Commensurate bond order}
\label{sec:A1}
In this section, we derive a set of self-consistent equations for the response of $P_\bk(\bq^\ast)$ to pointlike impurities for the case of  commensurate period-$m$ bond order.  In this case, the ordering wavevector $\bq^\ast$ satisfies  $m\bq^\ast = {\bf K}$, where ${\bf K}$ is a reciprocal lattice vector of the original lattice.

The  potential energy of electrons interacting with the  impurities is given by Eq.~(\ref{eq:V0}).
 This can be re-written as
\begin{equation}
\hat V = \sum_{\bk, \bk^\prime \in \mathrm{BZ}^\prime} \sum_\sigma \Psi^\dagger_{\bk\sigma} {\bf V}(\bk-\bk^\prime) \Psi_{\bk^\prime\sigma},
\label{eq:Vk1k2}
\end{equation}
where $\Psi_{\bk\sigma}$ is the column vector defined in Eq.~(\ref{eq:Psidef}),
${\bf V}(\bk-\bk^\prime)$ has matrix elements
\begin{equation}
V_{ab}(\bk-\bk^\prime)  =  V_i \frac{1}{N} \sum_I e^{-i (\bk+a \bq -\bk^\prime - b \bq)\cdot \bR_I},
\end{equation}
 $\bk$ and $\bk^\prime$ are now restricted to the reduced Brillouin zone, and $a,b \in [1,m]$.

\begin{figure}
\includegraphics[width=\columnwidth]{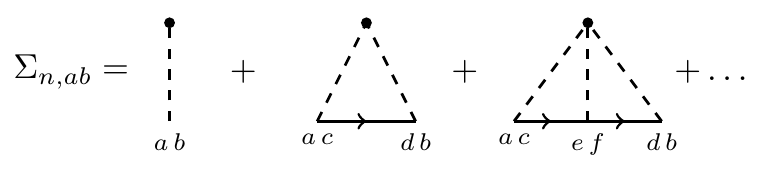}
\caption{Diagrams used in the SCTMA calculation of the self-energy matrix element $\Sigma_{n,ab}$.  Dashed lines represent scattering by the impurity potential $V_i$,  solid lines are the Green's function, and indices $a$, $b$, \ldots label reduced Brillouin zones for the initial and final momentum of the scattered particle. Internal indices are summed over.  These diagrams include scattering to all orders from a single impurity.  In the SCTMA, the Green's function contains the self energy due to impurity scattering, and is calculated self-consistently.}
\label{fig:SCTMA}
\end{figure}

The self energy is obtained from the sum of non crossing irreducible diagrams shown in Fig.~\ref{fig:SCTMA}.  These include all diagrams to linear order in $n_i$ due to scattering from the impurity potential: ${\bf \Sigma}_n = {\bf \Sigma}^1_n + {\bf \Sigma}^2_n +\ldots$, where ${\bf \Sigma}^j_n$ is $j$th order in $V_i$ and the subsript $n$ indicates that the self energy is evaluated at Matsubara frequency $\omega_n$.   As a result of disorder averaging, all terms depend on $\bk$ and $\bk^\prime$ only through a term $\delta_{\bk,\bk^\prime}$ that conserves momentum.  
 
 The first order term is 
\begin{eqnarray}
\Sigma^1_{ab}(\bk,\bk^\prime;i\omega_n) &=& \langle V_{ab}(\bk-\bk^\prime) \rangle \nonumber \\
&=& V_i \frac{1}{N} \sum_I  \frac{1}{N} \sum_{R_I} e^{-i (\bk+a \bq -\bk^\prime - b \bq)\cdot \bR_I}  \nonumber\\
&=&  n_i V_i \delta_{a,b} \delta_{\bk,\bk^\prime } \nonumber \\
&=& \Sigma^1_{n,ab} \delta_{\bk,\bk^\prime}
\end{eqnarray}
where $\langle \ldots \rangle = \frac{1}{N} \sum_{\bR_I} [\ldots]$  is the average over all possible positions for the $I$th impurity.   

Similarly, the irreducible second order term is 
\begin{eqnarray}
\Sigma^2_{n,ab} \delta_{\bk,\bk^\prime} &=&\sum_{\bk^\dprime} \sum_{cd} \langle V_{ac}(\bk-\bk^\dprime) G_{cd}(\bk^\dprime) V_{db}(\bk^\dprime-\bk^\prime) \rangle \nonumber \\
&=& n_i V_i^2 \delta_{\bk,\bk^\prime}  {\cal G}_{n,ab}
\end{eqnarray}
where 
\begin{equation}
{\cal G}_{n,ab} = \frac {1}{N^\prime} \sum_{\bk^\dprime\in \mathrm{BZ}^\prime} \frac 1m \sum_{c,d=1}^m G_{cd}(\bk^\dprime,i\omega_n) \delta_{a-b,c-d}, 
\label{eq:Gab}
\end{equation}
and where it is understood throughout this appendix that the Kronecker delta function $\delta_{a-b,c-d}$ is satisfied modulo $m$.  We have also explicitly written $N=N^\prime m$, where $N^\prime$ is the number of $\bk$-points in a single reduced Brillouin zone.  For reference,  
\begin{equation}
{\cal G}_{n,aa} \rightarrow {\cal G}^0_n; \quad {\cal G}_{n,a+1\,a} \rightarrow {\cal G}^{\bq^\ast}_n,
\end{equation}
in the limit of small $P_\bk(\bq^\ast)$, where $ {\cal G}^0_n$ and $ {\cal G}^{\bq^\ast}_n$ are defined in Appendix~\ref{sec:A2}.

The irreducible third order  term is
\begin{eqnarray}
\Sigma^{3}_{n,ab}  &=& n_i V_i^3  \frac {1}{N^\prime m} \sum_{\bk^\prime \in \mathrm{BZ^\prime} } \sum_{c,d=1}^m G_{cd}(\bk^\prime,\omega) \nonumber \\
&&\times \frac {1}{N^\prime m} \sum_{\bk^\dprime \in \mathrm{BZ^\prime}} \sum_{f,g=1}^m G_{fg}(\bk^\dprime,\omega)
\delta_{a-c+d-b,f-g} \nonumber \\
&=& n_i V_i^3 \sum_{h=1}^m {\cal G}_{n,ah} {\cal G}_{n,hb}.
\end{eqnarray}

At this point, the pattern is established:  the $j$th order term in the series is a matrix product of $j$ factors of the $m\times m$ matrix ${\cal G}_n$.  We define a t-matrix 
\begin{eqnarray}
{\bf T}_n & =& V_i {\bf 1} + V_i^2 {\cal G}_n + V_i^3 {\cal G}_n {\cal G}_n  + \ldots \nonumber \\
& =& V_i {\bf 1} + V_i {\cal G}_n{\bf T}_n \nonumber \\
&=& [{\bf 1}-V_i{\cal G}_n]^{-1} V_i,
\label{eq:Tmat}
\end{eqnarray}
where $[\ldots]^{-1}$ indicates a matrix inverse.
Then the  self-energy matrix is 
\begin{equation}
{\bf \Sigma}_n = n_i {\bf T}_n = n_iV_i[{\bf 1}-V_i{\cal G}_n]^{-1}.
\label{eq:Sigma}
\end{equation}

To determine the effect of impurity scattering on bond order,  one must simultaneously solve   Eq.~(\ref{eq:Sigma}) for the self energy and Eq.~(\ref{eq:Pdq}) for the order parameter.  These equations are linked by Eq.~(\ref{eq:fullG}) for the Green's function.

\section{Impurities in the Superconducting Phase}
\label{sec:B}
\subsection{$T_c$ equations for dirty superconductors}
\label{sec:B1}
We briefly review the $T_c$ equations for dirty superconductors in the absence of bond order, calculated with the SCTMA. Much of this discussion can be found elsewhere\cite{Schmitt-Rink:1986,Hirschfeld:1988}  and we include it here for completeness.  In the absence of bond order, Cooper pairs have zero center-of-mass momentum and the Hamiltonian is
\begin{equation}
\hat H = \sum_\bk \left [c^\dagger_{\bk\uparrow}, c_{-\bk\downarrow} \right ]
\left [ 
\begin{array}{cc}
\epsilon_\bk & \Delta_\bk \\ \Delta_\bk & -\epsilon_{-\bk}
\end{array} \right ] 
\left [ \begin{array}{c} c_{\bk\uparrow} \\ c^\dagger_{-\bk\downarrow} \end{array} \right ].
\end{equation}
Because of the particle-hole transformation for the spin-down component, the impurity potential is
\begin{equation}
\hat V = \frac{V_i}{N} \sum_{\bk,\bk^\prime} \left [c^\dagger_{\bk\uparrow}, c_{-\bk\downarrow} \right ]
\sum_I e^{-i(\bk -\bk^\prime)\cdot \bR_I} \otimes \tau_3 \left [ \begin{array}{c} c_{\bk^\prime \uparrow} \\ c^\dagger_{-\bk^\prime \downarrow} \end{array} \right ]
\end{equation}
where $\tau_3$ is a Pauli matrix in particle-hole space.  The impurity self-energy given by summing the SCTMA diagrams shown in Fig.~\ref{fig:SCTMA} is then
\begin{eqnarray}
\mathbf{\Sigma}_n &=& n_i \left[ V_i\tau_3 + V_i^2 \tau_3 {\cal G}(i\omega_n) \tau_3 + \ldots \right ] \nonumber \\
&=& n_i V_i \tau_3 \left [ 1-V_i {\cal G}(i\omega_n) \tau_3 \right ]^{-1} \\
&=& \Sigma^0_n \tau^0 + \Sigma^1_n\tau_1 + \Sigma^3_n\tau_3,
\end{eqnarray}
where 
\begin{eqnarray}
{\cal G}(i\omega_n) &=& -\frac{1}{N}\sum_{\bk} \frac{ \left [ \begin{array}{cc}
i\tilde \omega_n+\tilde \epsilon_\bk &
\tilde \Delta_\bk\\
\tilde \Delta_\bk &
i\tilde \omega_n-\tilde \epsilon_\bk 
\end{array}\right ] }{\tilde \omega^2_n+\tilde \epsilon_\bk^2} 
\label{eq:GSC} \\
&=&{\cal G}^0_n \tau_0 + {\cal G}^1_n \tau_1 + {\cal G}^3_n \tau_3,
\end{eqnarray}
and
\begin{eqnarray}
i\tilde \omega_n&=& i\omega_n - \Sigma^0_n = i\omega_n -  \frac{ n_iV_i^2 {\cal G}^0_n }{(1-V_i{\cal G}^3_n)^2 -  \left( V_i {\cal G}^0_n \right )^2 } \label{eq:omegatilde} \\
\tilde \Delta_\bk &=& \Delta_\bk + \Sigma^1_n  = \Delta_\bk  -  \frac{ n_iV_i^2 {\cal G}^1_n }{(1-V_i{\cal G}^3_n)^2  - \left( V_i {\cal G}^0_n \right )^2 } \label{eq:Dtilde} \\
\tilde \epsilon_\bk &=& \epsilon_\bk + \Sigma^3_n = \epsilon_\bk +  \frac{ n_iV_i \left( 1 - V_i {\cal G}^3_n\right ) }{(1-V_i{\cal G}^3_n)^2  - \left( V_i {\cal G}^0_n \right )^2 }
\end{eqnarray}
From the structure of Eq.~(\ref{eq:GSC}), one sees that ${\cal G}^0_n$ is pure imaginary, while ${\cal G}^1_n$ and ${\cal G}^3_n$ are real.
Equation (\ref{eq:GSC}) neglects terms of order $\tilde \Delta_\bk^2$, as these are small near $T_c$.
 $T_c$ is then obtained by solving the linearized equation 
\begin{eqnarray}
\Delta_\bk &=& -\frac{J_\mathrm{sc}}{N}  \sum_{\bk^\prime} g_\bk g_{\bk^\prime} \langle c_{-\bk^\prime \downarrow} c^\dagger_{\bk^\prime\uparrow} \rangle
\nonumber \\
&=& \frac{J_\mathrm{sc}T}{N}  \sum_{\bk^\prime,n} g_\bk g_{\bk^\prime} \frac{\tilde \Delta_{\bk^\prime }}{\tilde \omega^2_n+\tilde \epsilon_{\bk^\prime}^2},
\label{eq:SCTMA_Tc}
\end{eqnarray}
where $g_\bk = 1$ for isotropic $s$-wave superconductors and $g_\bk = \cos k_x - \cos k_y$ for $d$-wave superconductors.
In this work, numerical results for $T_c$ without bond order are generated by solving Eq.~(\ref{eq:SCTMA_Tc}) self-consistently.

To illustrate the role of each component of the self-energy,  and in particular the anomalous self-energy $\Sigma^1_n$, we take the simple case of a band with a constant density of states $N_0$. The components of ${\cal G}(i\omega_n)$ are 
\begin{eqnarray}
{\cal G}^0_n &=& -\frac{1}{N} \sum_\bk \frac{i\tilde \omega_n}{\tilde \omega^2_n+\tilde \epsilon_\bk^2 } \nonumber  \\
&=&-N_0 \int d\tilde \epsilon  \frac{i\tilde \omega_n}{\tilde \omega^2_n+\tilde \epsilon^2 } \nonumber \\ 
&=& -i\pi N_0 \mbox{sgn}(\omega_n)  \\
{\cal G}^3_n &=& -N_0 \int d\tilde \epsilon  \frac{\tilde \epsilon}{\tilde \omega^2_n+ \tilde \epsilon^2 } =0.
\end{eqnarray}
 It then follows that
\begin{equation}
\tilde \omega_n = \omega_n + \gamma\mbox{sgn}(\omega_n);\quad
\gamma = \frac{n_i \pi N_0 V_i^2}{1 + (\pi N_0 V_i)^2}
\label{eq:gamma}
\end{equation}
and 
\begin{equation}
\tilde \epsilon_\bk = \epsilon_\bk + \Delta \mu;
\quad 
\Delta \mu = \frac{n_i V_i}{1 + (\pi N_0 V_i)^2}.
\end{equation}

For $d$-wave superconductors 
\begin{equation}
{\cal G}^1_n = -\frac{1}{N} \sum_{\bk}\frac{\tilde \Delta_\bk}{\tilde \omega^2_n+\tilde \epsilon_\bk^2} = 0,
\end{equation}
so $\Sigma^1_n = 0$ and $\tilde \Delta_\bk = \Delta_\bk = \Delta g_\bk$.  Then, Eq.~(\ref{eq:SCTMA_Tc}) becomes
\begin{equation}
1 = \frac{J_\mathrm{sc} T}{N} \sum_{\bk,n} \frac{ g_\bk^2 }{\tilde \omega_n^2 + \epsilon_\bk^2}.
\label{eq:DSC}
\end{equation}
Because $\tilde \omega_n^2  = (|\omega_n|+\gamma)^2$,  the effect of $\gamma$ is to renormalize the Matsubara frequencies away from zero, which is qualitatively similar to raising the temperature in Eq.~(\ref{eq:DSC}).  Impurities thus impede $d$-wave superconductivity.

For isotropic $s$-wave superconductors $\tilde \Delta_\bk = \tilde \Delta$ and
\begin{equation}
{\cal G}^1_n = -\frac{1}{N} \sum_{\bk}\frac{\tilde \Delta}{\tilde \omega^2_n+\tilde \epsilon_\bk^2} = -N_0 \pi \frac{\tilde \Delta}{|\tilde \omega_n|} \Theta(\Lambda-|\tilde \omega_n|),
\label{eq:G1}
\end{equation}
where $\Theta(x)$ is a step function and $\Lambda$ is a cutoff that is typically of order the Debye frequency.
Then, combining Eq.~(\ref{eq:G1}), Eq.~(\ref{eq:Dtilde}) and Eq.~(\ref{eq:gamma}), we obtain the self-consistent equation
\begin{equation}
 \tilde \Delta = \Delta  + \gamma \frac{\tilde \Delta}{|\tilde \omega_n|},  \qquad (|\tilde \omega_n| < \Lambda)
\end{equation}
which has the solution 
\begin{equation}
\frac{\tilde \Delta}{|\tilde \omega_n|} = \frac{\Delta}{|\omega_n|}.
\label{eq:Hah}
\end{equation}
This result is directly relevant to the $T_c$ equation, which in this instance is given by Eq.~(\ref{eq:SCTMA_Tc}) with $g_\bk= 1$:
\begin{eqnarray}
\Delta &=& N_0 J_\mathrm{sc}T \sum_{|\tilde \omega_n|<\Lambda} \int  d\tilde \epsilon \frac{\tilde \Delta}{\tilde \omega^2_n+\tilde \epsilon^2} \nonumber \\
&=& \pi N_0 J_\mathrm{sc}T \sum_{|\tilde \omega_n|<\Lambda} \frac{\tilde \Delta}{| \tilde \omega_n |} \nonumber \\
&=& \pi N_0 J_\mathrm{sc}T \sum_{| \omega_n|<\Lambda} \frac{\Delta}{| \omega_n |}, 
\label{eq:Tcswave}
\end{eqnarray}
The last equality  follows from Eq.~(\ref{eq:Hah}), and the switch of the constraint from $|\tilde \omega_n|<\Lambda$ to $|\omega_n|<\Lambda$ introduces an error $\sim O(\gamma/\Lambda)$.  The key point of this derivation is that the the anomalous impurity self-energy $\Sigma^1_n$, which renormalizes $\Delta$, cancels the renormalization of $\omega_n$ by $\Sigma^0_n$, so that the $T_c$ equation is the same as in the clean limit.  In the $d$-wave case, where $\Sigma^1_n =0$, $T_c$ is reduced by impurities.

\subsection{$T_c$ in the bond ordered phase}
\label{sec:B2}
In this section, we derive the linearized self-consistent equation for the superconducting order parameter $\Delta_\bk(\bq)$ in the bond ordered phase.
From Eq.~(\ref{eq:SCEq}), we have
\begin{eqnarray}
\Delta^\alpha(a \bq^\ast) &=& -\frac{J_\mathrm{sc} }{mN^\prime} \sum_{\bk\in \mathrm{BZ}^\prime} \sum_{\ell=1}^m \eta^\alpha_{\bk_\ell} \langle c_{-\bk_\ell+a\bq^\ast\downarrow} c_{\bk_\ell \uparrow}\rangle \nonumber \\
&=& -\frac{J_\mathrm{sc}}{mN^\prime} \sum_{\bk\in \mathrm{BZ}^\prime} \sum_{\ell=1}^m \eta^\alpha_{\bk_\ell} \left [ {\cal F}_\bk \right ]_{\ell, \ell-a}
\label{eq:B1}
\end{eqnarray}
where ${\cal F}_\bk$ is the $m\times m$ anomalous Green's function with matrix elements
\begin{equation}
[{\cal F}_\bk ]_{ab} = -\langle c_{\bk_a \uparrow} c_{-\bk_b \downarrow} \rangle,
\end{equation}
and where it is understood that $\ell-a$ is modulo $m$.  To obtain ${\cal F}_\bk$, we solve the equations of motion:
\begin{widetext}
\begin{equation}
 \left [ \begin{array}{cc}
 i\omega_n - {\bf H}_\bk(\bq^\ast) - {\bf \Sigma}_n & -{\bf \Delta}_\bk - {\bf \tilde \Sigma_n} \\
  -{\bf \Delta}^\dagger_\bk - {\bf \tilde \Sigma_n} &  i\omega_n + {\bf H}_{-\bk}(-\bq^\ast)^T - \overline{\bf \Sigma}_n
 \end{array}\right ] 
 \left [ \begin{array}{cc} 
 { G} & {\cal F} \\
 \overline{\cal F} & \overline{G} \end{array} \right ]
 = \left [ \begin{array}{cc} {\bf 1} & 0 \\ 0 & {\bf 1} \end{array}\right ],
 \label{eq:EOM}
\end{equation}
to linear order in ${\bf \Delta}_\bk$.  To simplify the calculations, we make the approximation that
${\bf \tilde \Sigma}_n =0$, which is strictly true for pure $d$-wave superconductors.  We find in our numerical solutions that the non-$d$-wave components induced by the charge order are typically an order of magnitude smaller than the $d$-wave components, so that this result remains approximately true.  Then, we obtain the $m\times m$ matrix
\begin{equation}
{\cal F}_\bk = \left [ i\omega_n {\bf 1} - {\bf H}_\bk(\bq^\ast) -{\bf \Sigma}_n \right ]^{-1}  \left [ {\bf \Delta}_\bk \right ]
\left [ i\omega_n {\bf 1} + {\bf H}_{-\bk}(-\bq^\ast)^T - {\bf \overline{\Sigma}}_n \right ]^{-1}.
\end{equation}
Combining this with Eq.~(\ref{eq:B1}), we obtain
\begin{equation}
\Delta^\alpha(a\bq^\ast) = -\frac{J_\mathrm{sc} T}{mN^\prime}\sum_n \sum_{\bk\in \mathrm{BZ}^\prime} \sum_{\ell,\ell^\prime=1}^m \sum_\beta \eta^\alpha_{\bk_\ell} \eta^\beta_{\bk_{\ell^\prime}}  
\left [ i\omega_n - {\bf H}_\bk(\bq^\ast) - {\bf \Sigma}_n \right ]^{-1}_{\ell \ell^\prime}
\left [ i\omega_n + {\bf H}_{-\bk}(-\bq^\ast)^T -{\bf \overline{ \Sigma}}_n\right ]^{-1}_{\ell^\prime-c, \ell-a}
 \Delta^\beta(c\bq^\ast).
\end{equation}
\end{widetext}
This is the result shown in Eq.~(\ref{eq:M}).  We show in Appendix~\ref{sec:B3}, that $\overline{\bf \Sigma}_n = -{\bf \Sigma}_n^\ast$.

\subsection{Impurities at $T_c$ in the bond ordered phase}
\label{sec:B3}

In the superconducting state, Eq.~(\ref{eq:Vk1k2}) gives the potential energy of the impurities in the spin-up block.  In the spin-down block, we make a particle-hole transformation and let $\bk \rightarrow-\bk$.  The particle-hole transformation introduces a minus sign, but leaves the form of the potential otherwise unchanged.  Then, combining both spin-up electrons and spin-down holes, we obtain
\begin{equation}
\hat V = \sum_{\bk, \bk^\prime \in \mathrm{BZ}^\prime} \tilde \Psi^\dagger_{\bk} 
\left [ \begin{array}{cc} {\bf V}(\bk-\bk^\prime)  & 0 \\ 
0 & -{\bf V}(\bk-\bk^\prime)^\ast 
\end{array} \right ] \tilde \Psi_{\bk^\prime},
\label{eq:SCVk1k2}
\end{equation}
where $\tilde \Psi_{\bk}$ is a rank-$2m$ array of particle/hole annihilation operators, defined in Eq.~(\ref{eq:Psitilde}).

Because superconductivity modifies ${\bf \Sigma}_n$ and $\overline{\bf \Sigma}_n$ at second order in $\Delta_\bk$, it is neglected in the linearized equations near $T_c$.  The equations for ${\bf \Sigma}_n$ and $\overline{\bf \Sigma}_n$ are thus obtained by setting ${\bf \Delta}_\bk ={\bf \tilde \Sigma}_n = 0$ in Eq.~(\ref{eq:EOM}) for the Green's functions, and then performing the   SCTMA sums shown in Fig.~\ref{fig:SCTMA}.  Because the particle and hole blocks are decoupled in both the Green's functions and the impurity potential,  ${\bf \Sigma}_n$ and $\overline{\bf \Sigma}_n$ can be evaluated independently.

To linear order in ${\bf \Delta}_\bk$,  ${\bf \Sigma}_n$ is given by Eq.~(\ref{eq:Sigma});   $\overline{\bf \Sigma}_n$ satisfies an equation at $T_c$ similar to Eq.~(\ref{eq:Sigma}), but with $V_i \rightarrow -V_i$ and ${\cal G}_{n,ab} \rightarrow \overline{\cal G}_{n,ab}$, where
\begin{eqnarray}
 \overline{\cal G}_{n,ab}
&=& 
 \frac {1}{mN^\prime} \sum_{\bk\in \mathrm{BZ^\prime}} \sum_{c,d=1}^m \overline{G}_{cd}(\bk;i\omega_n) \delta_{a-b,c-d} \nonumber \\
&=& \frac {1}{mN^\prime} \sum_{\bk\in \mathrm{BZ}^\prime} \sum_{c,d=1}^m \left [ i\omega_n + {\bf H}_{-\bk}(-\bq)^T - \overline{\bf \Sigma}_n
\right ]^{-1}_{cd}  \nonumber \\ 
&&\times \delta_{a-c,b-d} 
\end{eqnarray}
Thus,
\begin{equation}
\overline {\bf \Sigma}_n(i\omega_n) = - n_iV_i \left [ 1 +  V_i \overline{\cal G}(i\omega_n)
\right ]^{-1}
\label{eq:B8}
\end{equation}
Because our solutions for $P_\bk(\bq)$ involve only $\cos(k_x+q_x/2)$ and $\cos(k_y+q_y/2)$,
it follows that ${\bf H}_{-\bk}(-\bq) = {\bf H}_{\bk}(\bq)$ and ${\bf H}_\bk(\bq)^T = {\bf H}_\bk(\bq)^\ast$
(ie.\ the matrix ${\bf H}_\bk(\bq)$ is Hermitian).  Then
\begin{eqnarray}
\overline{\cal G}_{n,ab} &=&  \frac{1}{mN^\prime} \sum_{\bk \in \mathrm{BZ}^\prime} \sum_{c,d} [i\omega_n + H_\bk(\bq)^\ast - \overline{\bf \Sigma}_n ]^{-1}_{cd}     \nonumber \\ 
&&\times \delta_{a-c,b-d}\nonumber \\
&=&  - \frac{1}{mN^\prime} \sum_{\bk\in \mathrm{BZ}^\prime } \sum_{c,d} { [i\omega_n - H_\bk(\bq) + \overline{\bf \Sigma}_n^\ast]^{-1}_{cd}}^\ast
  \nonumber \\ 
&&\times \delta_{a-c,b-d} 
\end{eqnarray}
Substituting this latter form into Eq.~(\ref{eq:B8}), it follows that $-\overline{\bf \Sigma}_n^\ast$ and
${\bf \Sigma}_n$ satisfy the same self-consistent equation.  We then make the identification
\begin{equation}
\overline{\bf \Sigma}_n = -{\bf \Sigma}_n^\ast.
\end{equation}

\bibliographystyle{apsrev} 
\bibliography{RPA}

\end{document}